\documentclass[usenatbib]{mn2e}

\usepackage{graphicx}
\usepackage{natbib}
\usepackage{rotating}
\usepackage[toc,page]{appendix}
\usepackage{aas_macros}
\usepackage{multirow}
\usepackage{placeins}
\usepackage{amssymb}
\usepackage{amsmath}
\usepackage{float}

\title[The X-ray Properties of Weak Lensing Selected Galaxy
  Clusters]{The X-ray Properties of Weak Lensing Selected Galaxy Clusters}
\author[P. A. Giles, B. J. Maughan, T. Hamana, S. Miyazaki,
  M. Birkinshaw, R.Ellis, R. Massey]{
\parbox[h]{\textwidth}{P. A. Giles$^1$\thanks{E-mail:
P.Giles@bristol.ac.uk}, B. J. Maughan,$^1$ T. Hamana$^2$,
  S. Miyazaki$^2$, M. Birkinshaw,$^1$ R. S. Ellis$^3$, R. Massey$^4$
}
\vspace*{12pt} \\
\parbox[h]{\textwidth}{
$^1$HH Wills Physics Laboratory, University of Bristol, Tyndall
  Avenue, Bristol, BS8 1TL, UK\\
$^2$National Astronomical Observatory of Japan, Mitaka, Tokyo
  181-8588, Japan\\ 
$^3$Astronomy Department, MS 249-17, California Institute of
  Technology, Pasadena, CA 91125, USA\\
$^4$Institute of Computational Cosmology, Durham University, South
  Road, Durham DH1 3LE, UK 
}}

\begin{document}

\date{18 February 2014}

\pagerange{\pageref{firstpage}--\pageref{lastpage}} \pubyear{2014}

\maketitle

\label{firstpage}

\begin{abstract}

We present the results of an X-ray follow-up campaign targeting 10
Weak Lensing (WL) selected galaxy clusters from a {\em Subaru} WL
survey.  Eight clusters were studied with dedicated {\em Chandra}
pointings, whereas archival X-ray data were used for the remaining two
clusters.  The WL clusters appear to fit the same scaling
relation between X-ray luminosity and temperature as X-ray selected
clusters.  However, when we consider the luminosity-mass relation, the
WL selected clusters appear underluminous by a factor 3.9$\pm$0.9 (or,
alternatively, more massive by 2.9$\pm$0.2), compared to X-ray
selected clusters.  Only by considering various observational
effects that could potentially bias WL masses, can this difference be
reconciled. We used X-ray imaging data to quantify the dynamical state
of the clusters and found that one of the clusters appears dynamically
relaxed, and two of the clusters host a cool core, consistent with
Sunyaev-Zel'dovich effect selected clusters.  Our results suggest that
regular, cool core clusters may be over-represented in X-ray selected
samples.  

\end{abstract}

\begin{keywords}
Galaxies: clusters: general -- Cosmology: observations -- X-rays: galaxies: clusters -- Gravitational lensing
\end{keywords}

\section{Introduction}
\label{sec:wlintro}

Clusters of galaxies are the largest gravitationally-collapsed
structures in the Universe. The study of their number density and
growth from high density perturbations in the early Universe
offers insight into the underlying cosmology 
\citep[e.g.,][]{2008MNRAS.387.1179M,2009ApJ...692.1060V}. However, the
use of clusters as a cosmological probe requires an efficient method to
find clusters over a wide redshift range, an observational method
of determining the cluster mass and a method to compute the
selection function or the survey volume in which clusters are found
\citep{2008MNRAS.387.1179M}. A variety of observational techniques has
been brought to bear on these  requirements, each with different
strengths and weaknesses. 

Early samples of clusters were based upon optical selection,
however these samples have traditionally suffered from
projection effects and uncertainties in the optical richness-mass
relation.  Recently, the first clusters have been detected in 
blind surveys using the Sunyaev-Zel'dovich effect 
\citep[SZE;][]{2009ApJ...701...32S,2010ApJ...722.1180V,2011A&A...536A...8P}. 
This method holds much promise due to the redshift independence of the 
SZE \citep{1999PhR...310...97B}, however the use of the
SZE as a mass estimator remains largely untested. Until now, the most
effective method of building large, well defined cluster samples has
been via X-ray selection. The high X-ray luminosities of clusters
makes it relatively easy to detect and study clusters out to high
redshifts, and X-ray cluster studies have provided a means to impose
tight constraints on cosmological parameters 
\citep[e.g.][]{2009ApJ...692.1060V,2010MNRAS.406.1759M}.

A common weakness of the three techniques outlined above is that the
clusters are selected based on the properties of their minority
baryonic content. However, cosmological constraints from those
clusters are based on their masses, dominated by dark matter. This
gives rise to complications both in the estimation of cluster masses
(which must be inferred from the observed baryon properties) and in
the determination of sample selection functions. Incorporating
selection functions in any cosmological model depends crucially on the 
form and scatter of the relationship between the observable used to
detect the cluster, and the cluster mass.

In principle, these complications may be avoided by defining cluster
samples through gravitational lensing, the most direct observational
probe of cluster masses. The development of weak lensing (abbreviated
WL throughout) techniques has enabled the detection of clusters via
the distortions of background galaxy shapes leading to the
construction of WL selected cluster surveys. These include the Deep
Lens Survey \citep{2006ApJ...643..128W} and our {\em Subaru} Weak Lensing Survey 
\citep{2007ApJ...669..714M}, both $\approx$20 deg$^2$, and the large
($\sim$170 deg$^{2}$) CFHT Legacy Survey
\citep{2012ApJ...748...56S}. However, while WL techniques are free
from assumptions about the relationship between
baryonic and dark matter in clusters, they are susceptible to the
possibility of projection of multiple structures along the line of
sight, leading to overestimates of cluster masses or false detections.

This paper aims to determine the X-ray properties of clusters detected
via their WL signal. By constructing scaling relations based on the
measured X-ray properties, we will determine whether the clusters
follow simple scaling theory, which is a key ingredient in the
determination of cosmological parameters.  One of the most important
measurements for use in cosmological studies is that of the cluster'
mass.  Much work has been done to determine cluster masses from weak
lensing observations, and the comparison to X-ray observables and
X-ray determined masses, however this is primarily based on X-ray
selected cluster samples
\citep[e.g.][]{2007MNRAS.379..317H,2010ApJ...721..875O,2011ApJ...737...59J,2012A&A...546A..79I,2012MNRAS.427.1298H,2013ApJ...767..116M,2014arXiv1402.3267I}. We
will compare the scaling relation between the X-ray
luminosity and WL mass from WL selected and X-ray selected
samples. The derivation of the WL cluster mass assumes
that the clusters follow spherical symmetry.  We investigate this
assumption and determine the dynamical state and cool-core fraction of
WL selected clusters.

The outline of this paper is as follows.  In $\S$~\ref{sec:wlreduc} we
discuss the sample selection and data reduction.  The derivation of
the cluster properties is given in $\S$~\ref{sec:wlprops}. In
$\S$~\ref{sec:wlresults} we present the results of our X-ray analysis and
derive scaling relations.  Our discussions and conclusions are
presented in $\S$~\ref{sec:wldisc} and $\S$~\ref{sec:wlconclusion}
respectively.  Throughout this paper we assume a cosmology with
$\Omega_{\rm M}$=0.3, $\Omega_{\Lambda}$=0.7 and 
$H_{0}$=70 km s$^{-1}$ Mpc$^{-1}$.

\section{Sample and Data Reduction}
\label{sec:wlreduc}

\begin{table*}
\caption[]{\small{Basic properties of the cluster sample. Columns: (1)
    = Source name; (2) = ObsID of the observation (4/5 digits {\em
      Chandra}, 10 digits {\em XMM}); (3) =
    Right Ascension at J2000 of the WL peak
    \citep{2009PASJ...61..833H}; (4) = Declination at J2000 of the
    WL peak \citep{2009PASJ...61..833H}; (5) = Redshift of
    the cluster as determined from the {\em Subaru} spectroscopic follow-up
    \citep{2009PASJ...61..833H}; (6) Cleaned exposure
    time; (7) Refers to the analysis method used to determine the
    properties of the cluster (see
    $\S$\ref{sec:wlprops}). $^{\dagger}$ cluster observed with {\em XMM}.}} 
\begin{center}
\begin{tabular}{ccccccc}
\hline\hline
Cluster & ObsID & RA & DEC & z & Exposure (ks) & Analysis \\
\hline 
SLJ0225.7--0312$^{\dagger}$ & 0553910201 & 02 25 43.2 & --03 12 36 &
0.1395 & 12 & Low SNR(b) \\
SLJ1647.7+3455$^{\dagger}$ & 0652400401,12303 & 16 47 47.5 & +34 55 13 & 0.2592 &
13 & Low SNR(b) \\ 
\hline
SLJ0850.5+4512 & 12305 & 08 50 31.7 & +45 12 12 & 0.1935 & 29 &
Standard \\
SLJ1000.7+0137 & 8022,8023,8555 & 10 00 45.5 & +01 39 26 & 0.2166 & 98
& Standard \\
SLJ1135.6+3009 & 12302 & 11 35 38.4 & +30 09 36 & 0.2078 & 11 & Low
SNR(c) \\
SLJ1204.4--0351 & 12304 & 12 04 22.9 & --03 50 55 & 0.2609 & 23 &
Standard \\
SLJ1335.7+3731 & 12307 & 13 35 45.6 & +37 31 48 & 0.4070 & 27 &
Standard \\
SLJ1337.7+3800 & 12306 & 13 37 43.7 & +38 00 57 & 0.1798 & 34 & Low
SNR(a) \\  
SLJ1602.8+4335 & 12308 & 16 02 52.8 & +43 35 24 & 0.4155 & 42 &
Standard \\
SLJ1634.1+5639 & 12309,13145 & 16 34 12.0 & +56 39 36 & 0.2377 &
48 & Low SNR(a) \\
\hline
\end{tabular}
\end{center}
\label{tab:wlclusts}
\end{table*}

Our clusters were selected from the {\em Subaru} WL survey of
\cite{2007ApJ...669..714M}.  The sample defined in this work was
constructed based upon confirmed cluster identification from 
spectroscopic follow up \citep{2009PASJ...61..833H}.  For 36 WL
cluster candidates 15-32
galaxy redshifts were obtained per cluster, with 28 candidates 
securely identified as clusters.  10 of these clusters (within the
redshift range 0.14$\leq${\em z}$\leq$0.42) were defined as a ``clean''
subset of clusters whose velocity dispersion could
be evaluated from at least 12 spectroscopic member galaxies, and whose
WL mass estimates are not affected by a neighboring system or field
boundary.  A summary of the 10 clusters investigated in this paper is
given in Table~\ref{tab:wlclusts}.  For eight of the clusters in the
sample we obtained dedicated {\em Chandra} pointings, whereas
SLJ1000.7 was observed in the COSMOS field with {\em Chandra} and
SLJ0225.7 was observed in the XMM\_LSS\_13 field with {\em
  XMM}.  We note however that the {\em Chandra} of SLJ1647.7 produced
low quality data, we therefore chose to use an archived {\em XMM}
observation for the analysis of this cluster.         

For the eight clusters analysed using {\em Chandra} observations we used the 
CIAO\footnote{See http://cxc.harvard.edu/ciao/} 4.4 software package 
with CALDB\footnote{See http://cxc.harvard.edu/caldb/} version 4.4.7
and followed standard reduction methods.  Since all observations were
telemetered in VFAINT mode additional background screening was  
applied\footnote{See
  http://cxc.harvard.edu/ciao/why/aciscleanvf.html}.  We inspected
background light curves of the observations following the 
recommendations given in \cite{2003ApJ...583...70M}, to search for 
possible background fluctuations.  None of the
observations were affected by periods of background flaring.  

In order to take the background into account, we
follow the method described in \cite{2006ApJ...640..691V}. Briefly,
{\em Chandra} blank-sky backgrounds were obtained, processed
identically to the cluster, and reprojected onto the sky to match the
cluster observation. We then renormalise the background in the 
9.5--12 keV band, where the {\em Chandra} effective area is nearly 
zero and the observed flux is due entirely to the particle background 
events.  Finally, to take into account differing contributions from
the soft X-ray background, the spectra were subtracted and
residuals were modeled in the 0.4-1keV band using an unabsorbed APEC
thermal plasma model \citep{2001ApJ...556L..91S} with a temperature of
0.18 keV. This model was included in the spectral fitting for the
cluster analysis.

The clusters SLJ0225.7 and SLJ1647 were observed by {\em XMM}. We used the
Science Analysis Software (SAS) version 12.0.1 and the most recent
calibration products as of Oct. 2013 for the analysis of these
clusters.    

Table~\ref{tab:wlclusts} lists the total cleaned exposure times for each
cluster. 

\section{Cluster Properties}
\label{sec:wlprops}

In this section we outline the methods used to determine the
cluster properties of our sample.  Where possible we use the method
outlined in Section~\ref{sec:wlyxprops}, however for clusters with low
SNR data, the properties are determined using one of the methods
described in Section~\ref{sec:wllowSNR}.  Table~\ref{tab:wlclusts}
lists the analysis method used for each cluster.

\subsection{The ``Standard'' Method}
\label{sec:wlyxprops}

To determine the cluster properties of our sample we followed the
procedures outlined in \cite{2012MNRAS.421.1583M}, for the {\em
  Chandra} observations. We detail our analysis for the {\em XMM}
observations of the clusters SLJ0225.7 and SLJ1647.7 in
Sect.~\ref{sec:wlslj0225} and \ref{sec:wlslj1647} respectively.
Briefly,  cluster spectra were extracted and fits performed in the 0.6
- 9.0 keV band with an absorbed APEC model, with the absorbing column
fixed at the Galactic value \citep{2005A&A...440..775K}.  We note that
due to the low SNR of five of the clusters (see~\ref{sec:wlnotes}),
the abundance was fixed at 0.3$Z_{\odot}$ throughout this analysis. We
determine the gas density profile for each cluster by converting the
observed surface brightness profile (constructed in the 0.7 -- 2.0 keV
band) into a projected emissivity profile, which is modeled by
projecting a density model along the line of sight.  We use the model
of 
\citet[][see that work for parameter definitions]{2006ApJ...640..691V}:
\begin{equation}
\hspace{1cm}
n_{\rm p}n_{\rm e} = \frac{n^2_0 (r/r_c)^{-\alpha}}{(1 +
  r^2/r^2_c)^{3\beta - \alpha/2}} \times (1 +
r^{\gamma}/r^{\gamma}_s)^{-\epsilon/\gamma}
\label{eq:wlgasdens}
\end{equation}
The same parameter constraints were employed as in
\cite{2006ApJ...640..691V} i.e. $\gamma$ is fixed at 3 and
$\epsilon<$5 to exclude nonphysical sharp density peaks. Gas masses
were determined from Monte Carlo realisations of the projected
emissivity profile based on the best fitting projected model
to the original data.  

The cluster temperature, gas mass and $r_{500}$ (the radius at which
the enclosed density of the cluster becomes 500 times the critical density at
the cluster's redshift) were determined through an iterative process.
We extract a spectrum from within an estimate of $r_{500}$ (with the
central 15\% excluded), integrate a gas density profile 
\citep[see][]{2008ApJS..174..117M} to determine the gas mass,
and then calculate a value for $Y_{\rm X}$ \citep[the product of the
temperature and gas mass;][]{2006ApJ...650..128K}.  A new $r_{500}$ was
then estimated from the $Y_{\rm X}$--M scaling relation of
\cite{2009ApJ...692.1033V},
\begin{equation}
\hspace{1cm}
M_{\rm 500} = E(z)^{-2/5}A_{\rm YM}\left(\frac{Y_{\rm
    X}}{3\times10^{14}M_{\odot} \rm keV}\right)^{B_{\rm YM}}
\end{equation}
with $A_{\rm YM}$ = 5.77$\times$10$^{14}$h$^{1/2}$M$_{\odot}$ and
$B_{\rm YM}$ = 0.57. Here, $M_{500}$ is the mass within $r_{500}$,
where $r_{500}$ is defined as $r_{500}$ =
(3$M_{500}$/4$\pi$500$\rho_{\rm c}$(z))$^{1/3}$, where $\rho_{\rm
  c}$(z) is the critical density of the universe at the cluster
redshift.  The $Y_{\rm X}$--$M_{500}$ relation assumes self-similar
evolution (corrected by $E(z)^{-2/5}$),
where E(z) = $\sqrt{\Omega_{\rm M}(1 + z)^3 + (1 - \Omega_{\rm M} -
  \Omega_{\Lambda})(1 + z)^2 + \Omega_{\Lambda}}$ 
\citep[justified by][]{2007ApJ...668..772M}.  An initial temperature
of 2 keV was used to determine an initial $r_{500}$ and then this
process was repeated until $r_{500}$ converged to within 1\%.  The
luminosity and temperature were measured from spectra extracted within
r$_{500}$ both with and without the central 15$\%$ of $r_{500}$
excluded.  Throughout we define $L_{\rm X}$ and kT as core excluded
cluster properties, and $L_{\rm X,c}$ and $kT_{\rm c}$ as core
included properties.  All luminosities quoted throughout are
bolometric unless otherwise stated.

\subsection{The ``Low SNR'' Method}
\label{sec:wllowSNR}

Due to the low SNR of the clusters resulting from the clusters being
less luminous than expected based on the WL mass (used when planning
the observations), five of the
clusters in our sample could not be analysed using the method
described in Sect.~\ref{sec:wlyxprops}.  We therefore used a variety of
methods in order to obtain the most accurate properties for these low
SNR clusters.  In order of preference, these were:
\begin{description}
\item[(a)] When the gas mass could not be reliably measured, we used
  the temperature alone to measure $r_{500}$.  We used the
  $r_{500}$-{\em T}
  relation given in \cite{2006ApJ...640..691V}, and followed the
  iterative process detained in Sect.~\ref{sec:wlyxprops} to determine
  the cluster properties within $r_{500}$.  We note that for
  consistency with the $r_{500}$-{\em T} relation, the central 70 kpc is
  excluded (instead of 0.15$\%r_{500}$ as in $\S$~\ref{sec:wlyxprops}).
\item[(b)] Extracting a spectrum within the highest SNR region for the
  cluster and determining the properties within this region.
  r$_{500}$ was then estimated using the $r_{500}$-{\em T} of
  \cite{2006ApJ...640..691V} and Lx was extrapolated to the radius by
  integrating under a $\beta$-profile with $r_{c}$=150 kpc and
  $\beta$=0.667. Again, the central 70 kpc is excluded for this analysis.  
\item[(c)] Using
  PIMMS\footnote{http://cxc.harvard.edu/toolkit/pimms.jsp} to
  determine the cluster luminosity from the count rate of the cluster
  observation when no spectroscopic analysis could be performed.  A
  global temperature of 2 keV was assumed for the cluster, and the count
  rate determined in an $r_{500}$ determined from the $r_{500}$-{\em T}
  relation of \cite{2006ApJ...640..691V}.
\end{description}

\subsection{Notes on Individual Clusters}
\label{sec:wlnotes}

Notes on the WL detections of the individual clusters can be
found in \cite{2009PASJ...61..833H}.  In this section we note any
peculiarities or points of interest for observations in which we
departed from analysis process described in Sect.~\ref{sec:wlyxprops}.
We note that five of the observations in our cluster sample were
analysed using the method described in Sect.~\ref{sec:wllowSNR}, and
these clusters are discussed below. 

\subsubsection{SLJ0225.7--0312}
\label{sec:wlslj0225}

The cluster SLJ0225.7--0312 had an archival {\em XMM}
observation taken in the XMM\_LSS\_13 field.  An image of the 
{\em XMM} observation is shown in Figure~\ref{fig:slj0225}(a).
The properties of the cluster were obtained following method (b)
outlined in Sect.~\ref{sec:wllowSNR}.  A spectrum was extracted
within [0-433]kpc for each of the {\em XMM} cameras and the
spectra fit simultaneously with the temperature of the hot APEC
components tied together. The net counts in the 0.3--7.0 keV band were
2635, 1019 and 1053 for the pn, mos1 (m1) and mos2 (m2) cameras
respectively, with SNR 40, 24 and 25 for the pn, m1 and m2 cameras
respectively.  Due to the large field of view of the {\em XMM}
cameras, a local background region was used for background
subtraction.  We found a temperature of {\em kT}=2.93$\pm$0.19 keV
for the cluster, corresponding to an $r_{500}$ of 798 kpc.   We found
an extrapolated luminosity within $r_{500}$ for the cluster of 
$L_{\rm X}$=(7.31$\pm$0.19)$\times$10$^{43}$ ergs s$^{-1}$.  
The analysis carried out above was also employed to determine the
properties of the cluster with the core included.  We derived a 
temperature of $kT_{\rm c}$=3.00$^{+0.18}_{-0.17}$ keV.  Extrapolating
outwards to $r_{500}$, we determined a luminosity of 
$L_{\rm X,c}$=(7.90$\pm$0.19)$\times$10$^{43}$ ergs s$^{-1}$.    

\subsubsection{SLJ1000.7+0137}
\label{sec:wlslj1000}

The cluster SLJ1000.7+0137 had three archival {\em Chandra}
observations taken in the COSMOS-6 field.
An image of the {\em Chandra} observation is shown in
Figure~\ref{fig:slj1000}(a).  The three individual observations were
analysed separately as described in Sect~\ref{sec:wlyxprops}.  The data
were then combined for certain stages of the analysis.  Source,
background and exposure maps were projected onto common coordinated
systems and summed. Source and background spectra were extracted for 
individual observations and fit simultaneously with the temperatures of
the hot APEC components tied together.  

\subsubsection{SLJ1135.6+3009}
\label{sec:wlslj1135}

In order to calculate the luminosity of this cluster, we followed
method (c) outlined in Sect.~\ref{sec:wllowSNR}.  We determined the
number of counts in a region centered on the location of the Brightest
Central Galaxy (BCG, see Figure~\ref{fig:slj1135} with the BCG
highlighted by the black circle), subtracting the number of counts
from the same region in a blank-sky background obtained for the
observation, scaled by the ratio of the source and background exposure
times.  The net number of counts obtained was 71$\pm$22 (significant
at the 3.2$\sigma$ level), in the 0.7--2.0 keV
band.  We calculated a bolometric luminosity of 
$L_{\rm X,c}$=(1.28$\pm$0.40)$\times$10$^{43}$ ergs s$^{-1}$ for the
cluster, with the errors derived assuming Poisson statistics.

\subsubsection{SLJ1337.7+3800}
\label{sec:wlslj1337}

To determine the cluster properties we used method (b) in
Sect.~\ref{sec:wllowSNR}, extracting spectra within a radius [70-215]kpc.
The net number of spectral counts, corrected for background, was then
130 (in the 0.6--9.0 keV band, with SNR=5.3).  We find a temperature
of {\em kT}=1.63$^{+0.68}_{-0.28}$ keV, corresponding to an $r_{500}$ of 574
kpc.  We derive an extrapolated bolometric luminosity of the cluster of 
$L_{\rm X}$=(6.58$\pm$1.31)$\times$10$^{42}$ ergs s$^{-1}$.  This
analysis was repeated including the core (obtaining 172 net
spectral counts in the 0.6--9.0 keV band, with SNR=6.45).  We determined a
temperature of $kT_{\rm c}$=1.63$^{+0.47}_{-0.23}$ keV and an extrapolated
bolometric luminosity of 
$L_{\rm X,c}$=(8.56$\pm$1.21)$\times$10$^{42}$ ergs s$^{-1}$. 

\subsubsection{SLJ1634.1+5639}
\label{sec:wlslj1634}

Due to observational constraints, the exposure was split into two
observations.  Therefore the analysis of the cluster was performed as
described in Sect.~\ref{sec:wlslj1000}.  The properties of this
cluster were obtained following the process outlined in method (a) in
Sect.~\ref{sec:wllowSNR}. We find a temperature of 
{\em kT}=1.37$^{+0.80}_{-0.44}$ keV and 
$L_{\rm X}$=(4.17$\pm$1.67)$\times$10$^{42}$ ergs s$^{-1}$.  Including
the core of the cluster we found a temperature and luminosity of
$kT_{\rm c}$=1.72$^{+0.95}_{-0.43}$ keV and
$L_{\rm X,c}$=(6.78$\pm$2.59)$\times$10$^{42}$ ergs s$^{-1}$.             

Substructure in SLJ1634 appears likely on the basis of the locations
of the galaxies, as marked in Figure~\ref{fig:slj1634}, which shows a
split in the population of galaxies in the northern and southern
region of the cluster. To check for the reality (or otherwise) of this
structure, a tree analysis of the locations of the galaxies in 3D
(position, velocity, redshift) space was undertaken.  The results
suggest a composite structure for the cluster, with the galaxies
separated into three separate groups.  The galaxies in
Figure~\ref{fig:slj1634} are colour coded with respect to each group
they are associated.  Group A (magenta squares), B (black squares)
and C (green squares) have a redshift of 0.231, 0.238 and  0.242
respectively. This composite structure may explain the 
large offset in the luminosity-mass relation for SLJ1634 when the WL mass is
calculated centered on the X-ray peak
(see Figs~\ref{fig:wlLM_bias},\ref{fig:wlLM_mahdavicomp}).

\subsubsection{SLJ1647.7+3455}
\label{sec:wlslj1647}

Although a dedicated {\em Chandra} observation was obtained for the
cluster SLJ1647.7+3455, due to the non-detection of the cluster, no
useful information could be determined.  We therefore used an archival
{\em XMM} observation to determine the properties of the cluster,
following the method outlined in Sect~\ref{sec:wllowSNR}(a).  The
  spectra were analysed in the same way as in
  Sect~\ref{sec:wlslj0225}. We obtained a temperature of
  {\em kT}=1.49$^{+0.11}_{-0.11}$ keV and
  $L_{\rm X}$=(1.38$\pm$0.12)$\times$10$^{42}$ ergs s$^{-1}$.
  Including the cluster core, we find 
$kT_{\rm c}$=1.58$^{+0.09}_{-0.10}$ keV and 
$L_{\rm X,c}$=(1.57$\pm$0.12)$\times$10$^{42}$ ergs s$^{-1}$.      

\section{Results}
\label{sec:wlresults}

In this section we present the results of our analysis of our 10
WL selected clusters.  We derive various scaling relations
for these clusters and compare to published results.  The measured
properties of the clusters are given in Table~\ref{tab:wlprops}, along
with the WL mass determined in \cite{2009PASJ...61..833H}
converted to $H_{0}$=70 km s$^{-1}$ Mpc$^{-1}$.  We note two
updates to the derivation of the WL masses given in
\cite{2009PASJ...61..833H}, and those presented in this work.  We use
the halo mass-concentration relation of \cite{2008MNRAS.390L..64D},
and the modified NFW profile given in \cite{2011MNRAS.414.1851O}

\begin{table*}
\caption[]{\small{Derived X-ray properties of the clusters with
    WL mass estimates (see $\S$~\ref{sec:wlresults}). Clusters in the
    top part of the table were analysed from {\em XMM} observations,
    in the bottom part from {\em Chandra}
    observations.} \label{tab:wlprops}} 
\begin{center}
\begin{tabular}{cccccccc}
\hline\hline
Cluster & E(z) & $L_{\rm X}$ & $L_{\rm X,c}$ & $L_{\rm
  0.1-2.4keV,X,c}$ & {\em kT} & $kT_{\rm
  c}$ & $M_{\rm WL,500}$ \\
 & & ($\times$10$^{43}$ ergs s$^{-1}$) & ($\times$10$^{43}$ ergs
s$^{-1}$) & ($\times$10$^{43}$ ergs s$^{-1}$) & keV & keV &
($\times$10$^{14}$M$_{\odot}$) \\ 
\hline 
SLJ0225.7--0312 & 1.070 & 7.31$\pm$0.19 & 7.90$\pm$0.19 &
 4.16$\pm$0.10 & 2.93$^{+0.19}_{-0.19}$ & 3.00$^{+0.18}_{-0.17}$ &
 1.97$^{+0.47}_{-0.47}$ \\
SLJ1647.7+3455 & 1.140 & 1.38$\pm$0.12 & 1.57$\pm$0.12 &
1.31$\pm$0.09 & 1.49$^{+0.11}_{-0.11}$ & 1.58$^{+0.09}_{-0.10}$ &
2.00$^{+0.67}_{-0.79}$ \\
\hline
SLJ0850.5+4512 & 1.100 & 0.78$\pm$0.14 & 0.92$\pm$0.13 &
0.83$\pm$0.13 & 1.21$^{+0.22}_{-0.16}$ & 1.16$^{+0.14}_{-0.15}$ &
1.09$^{+0.39}_{-0.43}$ \\
SLJ1000.7+0137 & 1.114 & 4.04$\pm$0.17 & 5.62$\pm$0.19 &
3.80$\pm$0.17 & 2.43$^{+0.29}_{-0.29}$ & 2.61$^{+0.41}_{-0.23}$ &
2.39$^{+0.46}_{-0.53}$ \\
SLJ1135.6+3009 & 1.108 & -- & 1.28$\pm$0.40 & 1.08$\pm$0.26 & -- & 
-- & 2.49$^{+0.50}_{-0.56}$ \\
SLJ1204.4--0351 & 1.141 & 3.43$\pm$0.35 & 3.97$\pm$0.38 &
2.92$\pm$0.28 & 2.16$^{+0.49}_{-0.37}$ & 2.16$^{+0.59}_{-0.22}$ &
1.20$^{+0.50}_{-0.60}$ \\
SLJ1335.7+3731 & 1.239 & 3.10$\pm$0.71 & 3.99$\pm$0.69 &
3.35$\pm$0.68 & 1.39$^{+0.63}_{-0.37}$ & 1.78$^{+0.79}_{-0.47}$ &
2.79$^{+0.90}_{-1.01}$ \\
SLJ1337.7+3800 & 1.092 & 0.66$\pm$0.13 & 0.86$\pm$0.12 &
0.37$\pm$0.05 & 1.63$^{+0.68}_{-0.28}$ & 1.63$^{+0.47}_{-0.23}$ &
1.24$^{+0.36}_{-0.39}$ \\   
SLJ1602.8+4335 & 1.245 & 12.7$\pm$1.14 & 16.4$\pm$1.22 &
8.50$\pm$0.54 & 3.91$^{+1.72}_{-0.88}$ & 4.45$^{+1.29}_{-1.12}$ &
2.66$^{+0.69}_{-0.71}$ \\
SLJ1634.1+5639 & 1.126 & 0.63$\pm$0.22 & 0.68$\pm$0.26 &
0.49$\pm$0.15 & 1.37$^{+0.77}_{-0.23}$ & 1.42$^{+0.83}_{-0.23}$ &
0.87$^{+0.39}_{-0.49}$ \\
\hline
\end{tabular}
\end{center}
\end{table*}

\subsection{The Luminosity-Temperature Relation}
\label{sec:wlLT}

\begin{figure}
\begin{center}
\includegraphics[clip=true, width=5.9cm, angle=270]{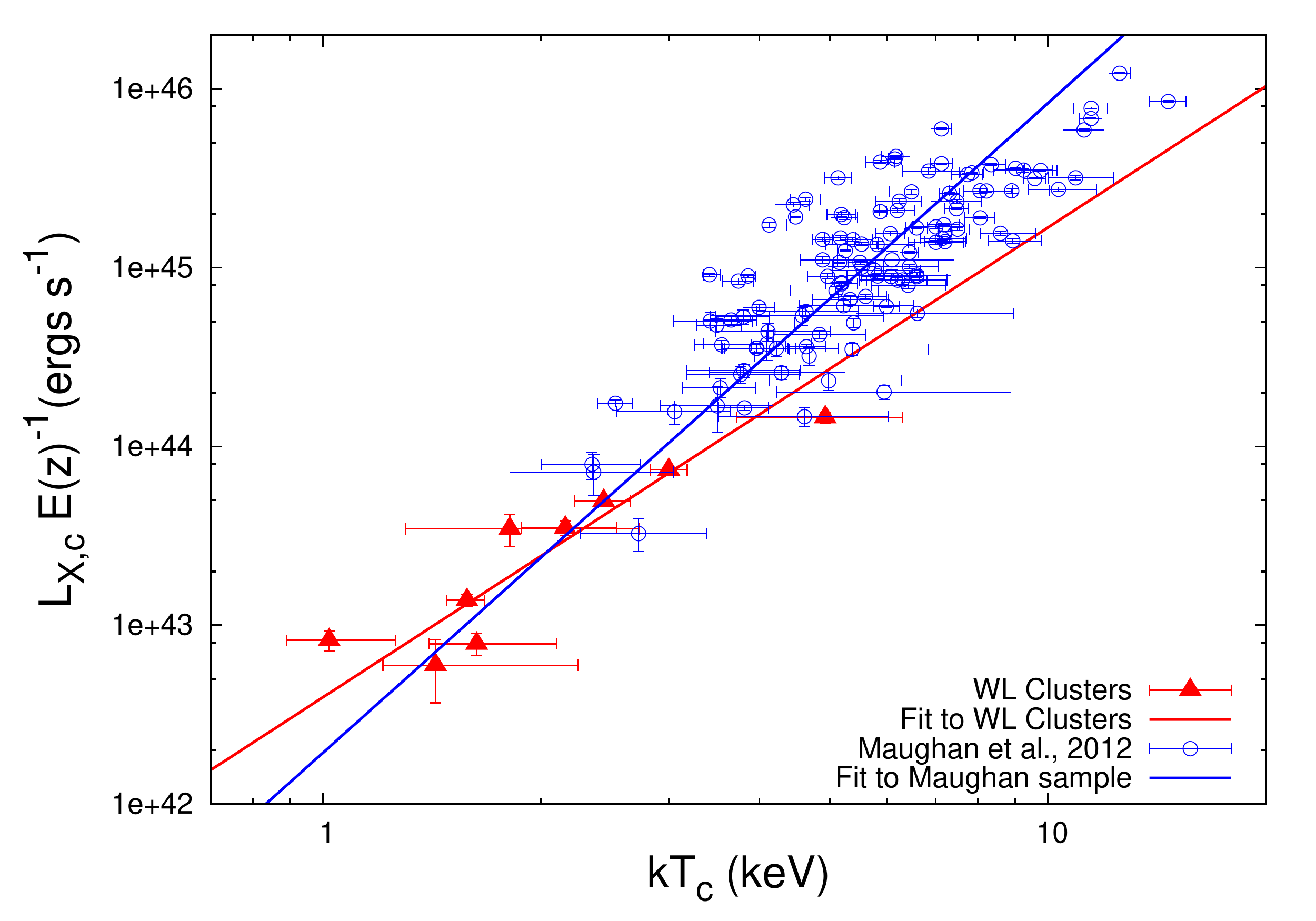}
\end{center}
\caption[]{\small{Figure showing the core included
    luminosity-temperature ($L_{\rm X,c}$-$kT_{\rm c}$) relation for
    the WL selected clusters (red circles).  For comparison
    we over-plot the $L_{\rm X,c}$-$kT_{\rm c}$ relation of a sample of
    114 X-ray selected clusters of \citet[][blue
    open circles]{2012MNRAS.421.1583M}. \label{fig:wlLTc}}}
\end{figure}

Here we derive the X-ray luminosity-temperature relation for our
clusters, omitting SLJ1135 for which a temperature could
not be obtained. Figure~\ref{fig:wlLTc} shows
the $L_{\rm X,c}$-$kT_{\rm c}$ relation for our clusters (red
triangles) compared with the LT relation given in
\citet[][blue open circles]{2012MNRAS.421.1583M}. The
\cite{2012MNRAS.421.1583M} sample of clusters was selected from all
available {\em Chandra} pointings as of 2006 November, correlated with
the NASA/IPAC Extragalctic Database (NED) and observations with a
galaxy group or cluster listed in NED within 10$^{\prime}$ of the {\em
Chandra} aimpoint were kept.  Further criteria included observations
only carried out with the ACIS-I detector, a galaxy group/cluster as
the target, and a published redshift $>$0.1. As the
\cite{2012MNRAS.421.1583M} sample of clusters covers
a wide redshift range  (0.1$<${\em z}$<$1.3), the luminosities of the
clusters were corrected for the expected self-similar evolution, given
by $L_{\rm X}\times$E(z)$^{-1}$.  The same correction was also
applied to our data. 

A power law of the form
$L_{X}=E(z)^{2}L_{0}(kT/kT_{0})^{B_{LT}}$ was fit to the data 
using the BCES orthogonal regression in log space
\citep{1996ApJ...470..706A} with $kT_{0}$ set at 2 keV.  From our fit
we find a normalisation of 
$L_{\rm 0}$=(2.44$\pm$0.51)$\times$10$^{43}$ ergs s$^{-1}$ and a slope of
$B_{\rm LT}$=2.63$\pm$0.69.  The slope and 
normalisation are consistent with the fit to the
\cite{2012MNRAS.421.1583M} sample, who find a normalisation of 
$L_{\rm 0,M12}$=(2.40$\pm$0.78)$\times$10$^{43}$ ergs s$^{-1}$ at 2
keV and a slope of $B_{\rm LT,M12}$=3.63$\pm$0.27 
(see Fig~\ref{fig:wlLTc}, blue solid line).  We conclude 
that the X-ray properties of the WL selected clusters scale
consistently with X-ray selected clusters in the
luminosity-temperature plane.        

\subsection{The Luminosity-Mass Relation}
\label{sec:wlLM}

\begin{figure*}
\begin{center}
\includegraphics[clip=true, width=11.0cm, angle=270]{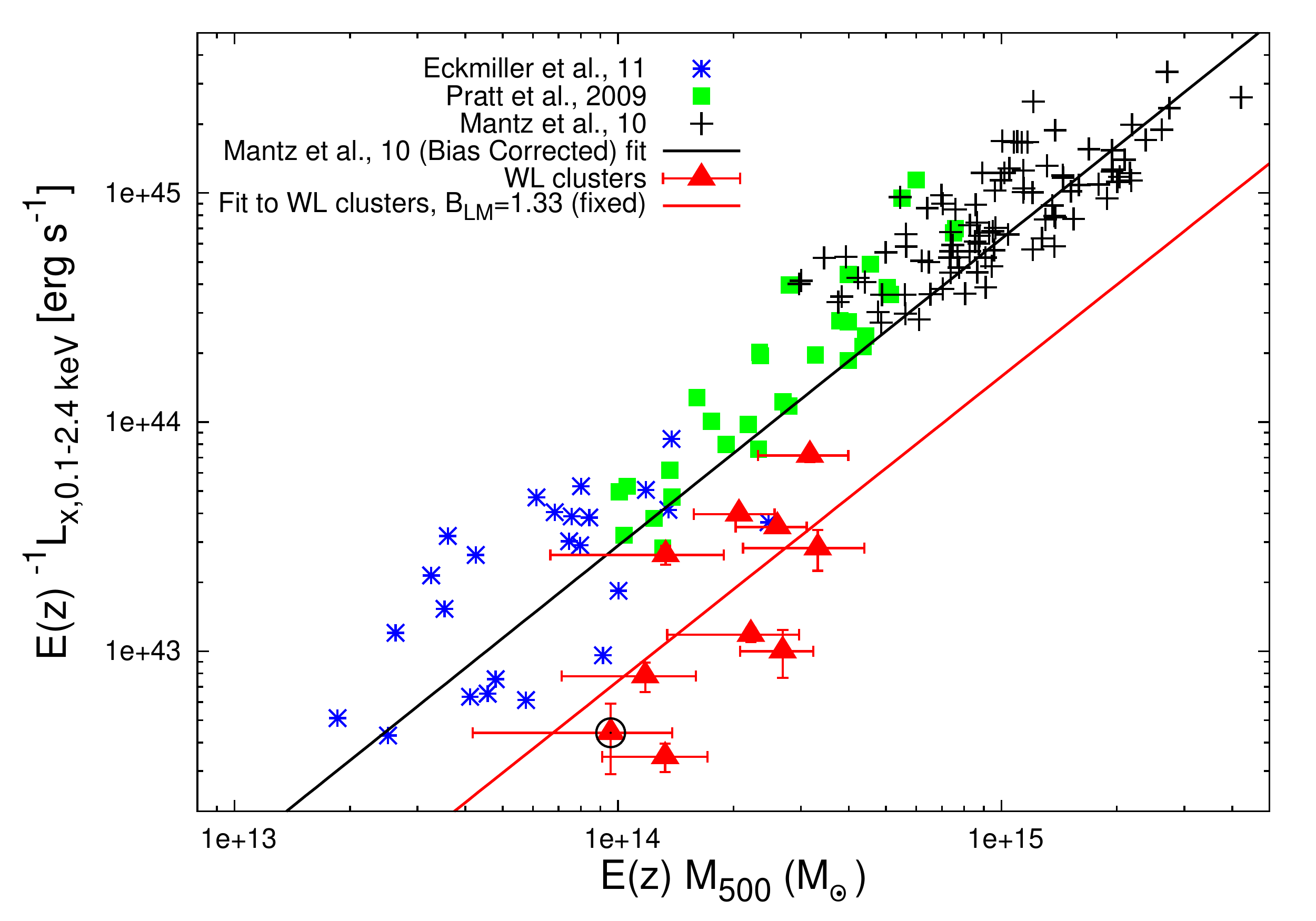}
\end{center}
\caption[]{\small{The luminosity-mass (LM) relation for
    our WL selected clusters (red triangles), whose masses
    are derived from WL. The cluster SLJ1634 is highlighted by the
    black circle.  For comparison we
    plot the data of X-ray selected clusters given in
    \citet[][black crosses]{2010MNRAS.406.1759M}, and the
    corresponding fit taking into account selection effects (black
    line).  Further comparisons are made to the REXCESS cluster sample
    \citep[][green squares]{2009A&A...498..361P} and galaxy group
    sample of \citet[][blue stars]{2011A&A...535A.105E}}\label{fig:wlLM}}
\end{figure*} 

Here we derive the luminosity-mass (LM) relation for our clusters,
with the masses estimated from the WL analysis presented in
\cite{2009PASJ...61..833H}, updated as described in
Section~\ref{sec:wlresults}.  Figure~\ref{fig:wlLM} shows
the luminosity-mass relation for our clusters (red triangles), and
compared to the data given in \citet[][black
 crosses]{2010MNRAS.406.1773M}.  Their clusters are taken from the
{\rm ROSAT} Brightest Clusters Sample
\citep[BCS;][]{1998MNRAS.301..881E}, the {\em ROSAT}-ESO Flux-Limited
X-ray Sample \citep[REFLEX;][]{2004A&A...425..367B}, and the bright
sub-sample of the Massive Cluster Survey \citep[Bright
  MACS;][]{2010MNRAS.407...83E}, all {\em z}$<$0.5. We note that
\cite{2010MNRAS.406.1773M} uses the gas mass as a robust, low-scatter
proxy for the total cluster mass.  The luminosities in
\cite{2010MNRAS.406.1773M} are calculated in the 0.1--2.4keV band
(cluster rest frame), we therefore calculate the luminosities of the
WL clusters in this band for the purpose of this comparison.  

Since we wish to compare weak-lensing
and X-ray selected clusters, we need to minimise any selection effects
in the X-ray comparison sample.   X-ray selected cluster samples
suffer from both Malmquist and Eddington biases. Malmquist bias refers
to the effect that a larger survey volume finds more luminous objects
for a given flux limit, and Eddington bias refers to the differential
scattering of objects across the detection threshold of a survey
due to the slope of the mass function.  The analysis presented in
\cite{2010MNRAS.406.1773M} takes fully into account these biases,
making it an ideal comparison for this work.  We also note that the
correction for the E(z) factor in \cite{2010MNRAS.406.1773M} for the
LM relation was based upon the best fit cosmology found in
\cite{2010MNRAS.406.1759M}.  For the purposes of the LM relation
comparison, the E(z) factor for each cluster is also calculated using
the \cite{2010MNRAS.406.1759M} cosmology.  On average, {\em E(z)} changes by
$\approx$3$\%$.    

A power law of the form
$E(z)^{-1}L_{\rm X}=L_{0}(E(z)M/M_{0})^{B_{\rm LM}}$ was fit to the WL clusters in
log space, with $M_{0}$ set at 2$\times$10$^{14}$M$_{\odot}$.  Due to the
low quality of the data and the small number of data points we fixed
the slope ($B_{\rm LM}$) of the relation at the self similar expectation of
1.33.  We note that since the slope is fixed, we perform a $\chi^{2}$
fit to the cluster LM relation.  For a mass of
2$\times$10$^{14}$M$_{\odot}$ we found $L_{0}$=(1.9$\pm$0.2)$\times$10$^{43}$
ergs s$^{-1}$ (see red solid line Figure~\ref{fig:wlLM}).  The fit from 
\cite{2010MNRAS.406.1759M} is represented by the solid black line.
The low normalisation compared to the data demonstrates the size of 
the Malmquist and Eddington biases (which were included in the
analysis of the \cite{2010MNRAS.406.1759M} data.
The \cite{2010MNRAS.406.1759M} LM relation at the same mass predicts
$L_{0}$=7.3$\times$10$^{43}$ ergs s$^{-1}$.  Compared with
X-ray selected clusters, X-ray luminosity is lower at a given mass by
a factor 3.9$\pm$0.9.  Conversely, $M_{\rm WL,500}$ is higher for a given
luminosity by a factor 2.9$\pm$0.2. We discuss the
nature of this large offset in the LM relation in
Sect.~\ref{sec:wldisc}.  

The main drawback in the comparison with the
\cite{2010MNRAS.406.1759M} sample of clusters is the mass range
covered by the respective cluster samples.  The
\cite{2010MNRAS.406.1759M} sample of clusters have masses
$M_{500}\ge$2.7$\times$10$^{14}$M$_{\odot}$, whereas our WL sample
are all $<$2.7$\times$10$^{14}$M$_{\odot}$.  To ensure the mass range
of the comparison sample is not playing a role in the large observed
offset, we compare to a sample of clusters probing the lower mass end
and a sample of galaxy groups.  We use the REXCESS
\citep{2007A&A...469..363B} sample of clusters studied in
\cite{2009A&A...498..361P} which span the mass range
$\approx$10$^{14}$-10$^{15}$M$_{\odot}$, and the sample of galaxy
groups presented in \cite{2011A&A...535A.105E} spanning the mass range
$\approx$0.05-2$\times$10$^{14}$M$_{\odot}$.  The masses of the 
\cite{2009A&A...498..361P} clusters were calculated using the 
$Y_{\rm X}$-Mass relation given in \cite{2007A&A...474L..37A}, and the
masses of the \cite{2011A&A...535A.105E} groups were calculated by
assuming hydrostatic equilibrium and spherical symmetry \citep[see
  Sect. 3.4 in][]{2011A&A...535A.105E}. In both cases the luminosities
are calculated in the 0.1-2.4 keV band (cluster rest frame). We plot
the clusters on the LM relation in Figure~\ref{fig:wlLM}, with the
\cite{2009A&A...498..361P} clusters given by the green squares and
the \cite{2011A&A...535A.105E} groups given by the blue stars.  A  
visual inspection shows that all the X-ray selected comparison groups
lie on approximately the same LM relation.  Although the
\cite{2009A&A...498..361P} and \cite{2011A&A...535A.105E} samples appear
to lie slightly above the \cite{2010MNRAS.406.1759M} relation, this is
consistent with a similar amount of bias as in the
\cite{2010MNRAS.406.1759M} sample, with the relation correcting for
these effects.  The offset of the WL selected clusters is
still evident down to these low masses.              

\subsection{X-ray Morphology of Weak Lensing Selected Clusters}
\label{sec:wlmorph}

\subsubsection{The dynamical state of weak lensing selected clusters}
\label{sec:wlstate}

We wish to investigate the dynamical state of our WL
selected clusters as compared to X-ray selected cluster samples.  
We use the centroid shift method of \cite{2006MNRAS.373..881P} to
determine the dynamical state of the cluster.  The centroid shift
($\langle${\em w}$\rangle$) was defined as the standard deviation of the
distance between the X-ray peak and centroid, measured within a series
of circular apertures centered on the X-ray peak and decreasing in
steps of 5$\%$ from $r_{500}$ to 0.05$r_{500}$. The
$\langle${\em w}$\rangle$ values are given in Table~\ref{tab:wldynamical}.  
\cite{2012MNRAS.421.1583M} used $\langle${\em w}$\rangle$ to distinguish
between ``relaxed'' and ``unrelaxed'' clusters.  For the
\cite{2012MNRAS.421.1583M} sample of 114 clusters they adopted a
value of $\langle${\em w}$\rangle$=0.006$r_{500}$ below which clusters are
classed as ``relaxed'', and above which are classed as ``unrelaxed''.
Using this threshold value, 25$\%$ of the \cite{2012MNRAS.421.1583M}
sample was classed as ``relaxed''.  If we adopt this value for our
sample of WL selected clusters, we find that none of the
clusters in Table~\ref{tab:wldynamical} qualify as ``relaxed''.

The value of $\langle${\em w}$\rangle$ is somewhat subjective, and the value
used to separate relaxed and unrelaxed clusters can depend on the
sample of clusters used, and indeed on each individual study.
\cite{2009A&A...498..361P} used the REXCESS sample of clusters and
defined relaxed/unrelaxed clusters using the same method of
calculating $\langle${\em w}$\rangle$ as we employ here.
\cite{2009A&A...498..361P} found that $\sim$60$\%$ of their clusters
has $\langle${\em w}$\rangle<$0.01, while for our WL clusters we find that
only one of the seven clusters with measurable $\langle$w$\rangle$
values are below this threshold.  We note however that this comparison
is only based on a small number of clusters from the WL.

This result appears analogous to the recent cluster sample from the
early release Planck all sky survey \citep{2011A&A...536A...9P}.
These clusters were detected via their SZ signals, and 25
candidates were followed up with snapshot {\em XMM} observations.
Through a visual inspection of the gas density profiles constructed
for each cluster, a large proportion of these SZ selected clusters
appear morphologically disturbed. Since the WL selected clusters
studied in our sample also appear to be morphologically disturbed,
selecting clusters from either their SZ or WL signal, as compared to
selecting cluster via their X-ray emission, appears to give a more
representative distribution of morphologies for clusters.  

\subsubsection{The cool-core fraction of weak lensing selected clusters}  
\label{sec:wlcoolcore}

The absence of relaxed clusters in our sample as indicated by
$\langle${\em w}$\rangle$ suggests we should also see few cool core (CC)
clusters.  This also has important
implications for cosmology as the CC fraction and its evolution may be
over-represented/biased in X-ray samples
\citep{2010A&A...521A..64S}. This result is not unexpected due to the
high X-ray surface brightness at the center of CC clusters, making them
easier to detect in X-ray cluster surveys.  \cite{2010A&A...513A..37H}
presented a comprehensive study of 16 CC diagnostics for the HIFLUGCS
sample of 64 clusters.  They found that, for clusters with low data
quality, the cuspiness of the gas density profile is the preferred
method of determine the presence of a CC.  Cuspiness is defined as the
logarithmic slope of the gas density profile at a radius of
0.04$r_{500}$, modeled using the gas density profiles given by
equation~\ref{eq:wlgasdens},  with the uncertainties
derived from the cuspiness values measured from Monte Carlo
realisations of the gas density profile.  We obtained the cuspiness
values from gas density profiles fitted to surface brightness profiles
derived from images binned by a factor 2, with each radial bin
containing at least 10 cluster counts.  The same fitting process was
followed as in $\S$~\ref{sec:wlyxprops} and the cuspiness values were
derived from these profiles.  Table~\ref{tab:wldynamical} lists the
values of the cuspiness for each cluster where a gas density profile
could be obtained.

\cite{2012MNRAS.421.1583M} also used the cuspiness to determine the
presence of a CC in their sample of 114 clusters.  Above a value of
0.8 clusters were considered to host a CC, and below they were not.
If we adopt this value for our sample of 6 clusters with measurable
cuspiness, we find that one of the clusters is considered to host a
CC.  We note that the \cite{2012MNRAS.421.1583M} sample spans a wide
range in redshift and data quality, and again the threshold in
cuspiness used to define CC/NCC clusters is subjective depending on
the sample being studied.  The
\cite{2010A&A...513A..37H} sample consists of clusters with very high
data quality and the value of the cuspiness is well
constrained for all clusters.  \cite{2010A&A...513A..37H} adopted a
value for cuspiness of 0.7 to segregate between CC and non-CC
clusters.  If we adopt this value of cuspiness for our sample of
clusters, we find only 2/6 of the clusters host a CC.  In both cases
we find a smaller percentage of CC clusters in comparison to these
samples of clusters, we find 16.6$\%$ of clusters host a CC when using
a cuspiness value of 0.8, as compared to 26.3$\%$ found in
\cite{2012MNRAS.421.1583M}.  We find that 33.3$\%$ of clusters host a
CC when using a cuspiness value of 0.7, as compared to 54.7$\%$ found
in \cite{2010A&A...513A..37H}. This suggests that CC clusters
may be over-represented in X-ray selected samples of clusters.          

\begin{table}
\caption[]{\small{Table of the derived dynamical properties of the
    cluster sample.}\label{tab:wldynamical}} 
\begin{center}
\begin{tabular}{cccc}
\hline\hline
Cluster & $\langle${\em w}$\rangle$(10$^{-3}r_{500}$) & cuspiness & $\epsilon_{\rm bcg}$ \\ 
\hline 
SLJ0225.7--0312 & -- & -- & 0.07 \\
SLJ1647.7+3455 & -- & -- & 0.15$\pm$0.01 \\
\hline
SLJ0850.5+4512 & 31$\pm$28 & 0.71$^{+0.44}_{-0.64}$ & 0.13$\pm$0.01\\
SLJ1000.7+0137 & 105$\pm$4 & 0.02$^{+0.15}_{-0.01}$ & 0.20$\pm$0.02 \\
SLJ1135.6+3009 & -- & -- & 0.40$\pm$0.01 \\
SLJ1204.4--0351 & 28$\pm$13 & 0.13$^{+0.55}_{-0.08}$ & 0.16$\pm$0.02 \\
SLJ1335.7+3731 & 12$\pm$8 & 0.24$^{+0.48}_{-0.17}$ & 0.11$\pm$0.02 \\
SLJ1337.7+3800 & 44$\pm$30 & 0.97$^{+0.19}_{-0.16}$ & 0.23$\pm$0.01 \\  
SLJ1602.8+4335 & 7$\pm$4 & 0.53$^{+0.17}_{-0.51}$ & 0.14$\pm$0.02 \\
SLJ1634.1+5639 & 87$\pm$85 & -- & 0.15$\pm$0.01 \\
\hline
\end{tabular}
\end{center}
\end{table}

\section{Discussion}
\label{sec:wldisc}

In Sect.~\ref{sec:wlLM} we showed that WL
selected galaxy clusters are underluminous by a factor $\approx$3.9
for a given mass, compared to X-ray selected clusters.  In terms of
the mass, we find that the WL selected clusters are over-massive by a
factor $\approx$3 compared to X-ray selected clusters. The consistency
of the $L_{\rm Xc}$-$kT_{\rm c}$ relation with X-ray selected clusters
(see Fig~\ref{fig:wlLTc}) suggests that the discrepancy may more
likely arise as a result of a bias in the mass.  We therefore first
investigate the nature of this offset through a study of systematic
effects on the WL mass estimates.          

\subsection{Possible Systematics in the WL Masses}
\label{sec:systematics}

In the following sections we present a series of plausible effects
that could bias the WL mass.  These biases are presented as a
cumulative effect on the WL mass, and we investigate the effect these
biases have on the LM relation. 

\subsubsection{The Effect of Cluster Centre Position on Weak Lensing Mass}
\label{sec:wlcentroid}
   
The X-ray analysis presented in this paper is independent of the WL
analysis given in \cite{2009PASJ...61..833H}, and as such, different
 locations of the cluster centre are used for the two analysis.
The WL masses given in Table~\ref{tab:wlprops} are derived from
fitting NFW models to shear profiles centered on the peak position in
the WL mass maps, whereas the luminosities are derived centered on the X-ray
centriod.  Here we investigate the effect of deriving the WL masses
from profiles centered on the X-ray centroid.  

\begin{figure}
\begin{center}
\includegraphics[clip=true, width=6.0cm, angle=270]{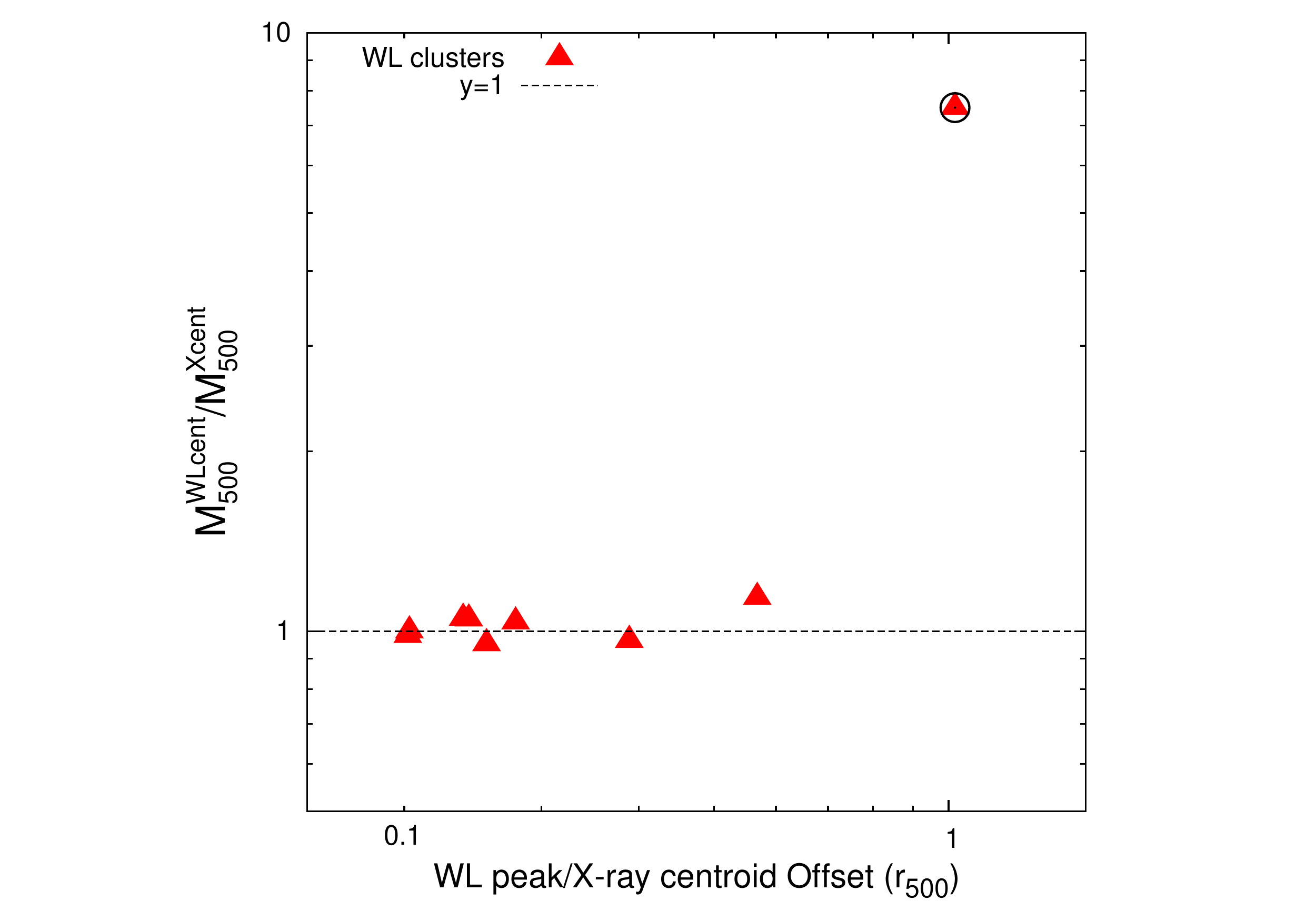}
\end{center}
\caption[]{\small{Plots the ratio of the WL mass derived
    within the X-ray $r_{500}$ at the WL peak and X-ray centroid,
    against the offset of the WL peak and X-ray centroid (plotted in
    units of $r_{500}$).  The cluster SLJ1634 is highlighted by the
    black circle.}\label{fig:centoffset}}
\end{figure}

Figure~\ref{fig:centoffset} shows the ratio of the WL mass centered on
the peak of the WL signal to the WL mass centered on the X-ray
centroid ($M^{\rm WLpeak}_{500}$/$M^{\rm Xcent}_{500}$) against
the offset of the WL peak and X-ray centroid (in units of r$_{500}$).
The cluster SLJ1634 is highlighted by the black circle.  This cluster
shows evidence for a composite structure (see $\S$\ref{sec:wlslj1634})
and the WL mass decreases by a factor $\sim$7 when centered at the
X-ray peak. We note for the cluster SLJ1135, the X-ray analysis was
centered on the WL peak due to the low SNR of the X-ray data, and no
X-ray centroid could be obtained.  It is thought that any offset
between the WL and X-ray would be minimal since the WL peak is aligned
closely with the BCG of the cluster (see Fig~\ref{fig:slj1135}(b)).
This figure shows that the ratio of the masses is consistent with a
ratio of 1.0 (given by the black dashed line).  This result is
unsurprising due to the radial range within which the WL masses are
derived (2$<$$\theta$$<$20$^{\prime}$), to neglect contamination from
cluster galaxies.  These results are consistent with the results from 
\cite{2010PASJ...62..811O}, who find that the relative inaccuracy of
cluster centroid position has a negligible effect on the
resulting $\chi^{2}$ value of the NFW fit to their distortion
profiles, and hence the mass derivation.                 

The effect of using WL masses centered on the X-ray centroid on the LM
relation is shown in Figure~\ref{fig:wlLM_bias}(dashed blue line).
The original LM relation from Figure~\ref{fig:wlLM} is given by the
dashed green line. By calculating the WL masses centered on the X-ray
centroids, the masses have decreased on average by a factor of $\approx$1.1.  

\subsubsection{''Deboosting`` the Weak Lensing Mass Estimates}
\label{sec:eddbias}

Here we correct the weak lensing masses by ``deboosting'' the shear
signal, which is analogous to the effect of Eddington bias (see
Sect~\ref{sec:wlLM}).  Since our sample is based upon a noisy
indicator of the cluster mass, the shear signal ($\kappa$), clusters
with high mass but a low observed $\kappa$ will not be included in our
sample, and conversely clusters with low masses but high observed
shear will be included in our sample.  Our cluster sample will
therefore be biased to clusters with a high shear
signal relative to their mass, and due to the slope of the mass
function, there will be more low mass than high mass clusters close to
our detection limit.  The overall effect will be to measure high mass
values, due to the high shear signal, and so bias the sample to higher
masses. An application of ``deboosting'' can be found in
\cite{2006MNRAS.372.1621C} for a SCUBA source catalog. 

In order to calculate ``deboosted'' mass estimates, we follow a
similar method to that carried out in \cite{2006MNRAS.372.1621C}.
Initially, the cluster mass is estimated from the shear profile ($\kappa$) via
$\chi^2$ fitting with an NFW model.  We also have a theoretical
prediction of the shear profile for a modified NFW profile.  We
compute the $\chi^2$ following the standard procedure from the
observed shear profile and the model prediction as a function of halo
mass M ($\chi^{M}$).  From $\chi^{M}$ we compute the best-fit
(non-``deboosted'') WL mass and likelihood interval.  Next we consider
$P={\rm exp}(-\chi^{2}/2)$ as $P(\kappa,\sigma\mid M)$, and consider the
halo mass function as p(M).  From Bayes' theorem we therefore have
$P(M\mid\kappa, \sigma) = P(M)P(\kappa, \sigma\mid M)$.  We then
compute $\chi^{2}_{\rm deboost}=-2{\rm ln}(P(M\mid\kappa, \sigma))$ and find
the best fit mass and the likelihood function. 

This correction decreases the WL mass by an average of
$\approx$1.27.  This is shown in Figure~\ref{fig:wlLM_bias} as
the dotted pink line, representing the cumulative effect of the
centroid position (see $\S$~\ref{sec:wlcentroid}) and Eddington bias
on the LM relation.

\subsubsection{Weak Lensing Biases due to Halos Elongated Along the Line of Sight}
\label{sec:wlellipt}

An additional bias involved in the estimation of weak lensing masses
is that due to a triaxial shape of the cold dark matter (CDM) halo 
\citep{1992ApJ...399..405W,2002ApJ...574..538J,2006ApJ...646..815S}.
This can lead to errors in the weak lensing masses by $\pm$(10-50)$\%$
when spherical symmetry is assumed
\citep{2007MNRAS.380..149C,2010A&A...514A..93M,2011ApJ...740...25B,2012MNRAS.421.1073B}.
In order to correct for the effect of a triaxial CDM halo we use the
set of simulated clusters studied in \cite{2010A&A...514A..93M}.
Figure 17 of \cite{2010A&A...514A..93M} shows the ratio of the
 weak+strong lensing mass (a nonparametric method combining both
strong and weak lensing regimes to calculate cluster mass) to true
cluster mass as a function of angle with the projection axis (the
angle between the major axis of the cluster inertia ellipsoid and the
axis along which the mass distribution is projected).  Since
our weak lensing masses are based upon NFW profiles, we combine Figure
17 with Figure 16(a) from \cite{2010A&A...514A..93M}, to derive the
same relation appropriate for NFW profile mass estimates. 

It is noted in \cite{2012ApJ...754..119M} that their plot of
fractional deviation in mass from a self-similar
$M_{\rm 500,WL}$-$Y_{\rm sph}$ as a function of brightest cluster galaxy (BCG)
ellipticity ($\epsilon_{\rm BCG}$) looks strikingly similar to Figure 17
 presented in \cite{2010A&A...514A..93M}.  They find that clusters with a BCG
ellipticity $\le$0.15 have the largest deviation in mass from the
self-similar relation.  By considering BCGs as prolate systems whose
major axis is aligned with the major axis of the CDM halo, circular
BCGs indicate the major axis is close to the line of sight (LOS) through the
cluster. The viewing angle ($\psi$) can be calculated as a function of
the observed axis ratio ({\em q}) and intrinsic BCG axis ratio ($\delta$), using   
\begin{equation}
\hspace{0.2cm}
\psi = {\rm arccos}\left(\sqrt\frac{1 - (\delta / q)^2}{1 -
  \beta^2}\right)
\label{equ:viewang}
\end{equation}
where $q = b/a$, and adopting $\delta=0.67$ \citep{2010MNRAS.404.1490F}.
A $\epsilon_{\rm BCG}$=0.15 corresponds to a viewing angle of
$\psi\simeq$34$^{\circ}$, we therefore use clusters below this angle from
\cite{2010A&A...514A..93M} to calculate the average of the weak
lensing mass to true mass and correct our masses by this factor.  

In order to correct our weak lensing masses, we calculate
$\epsilon_{\rm BCG}$ for each cluster from the {\em Subaru} images.
For clusters which have $\epsilon_{\rm BCG}\le$0.15 (i.e. with a viewing
angle $\le$34$^{\circ}$), we divide the corresponding WL mass by 1.17,
and clusters with $\epsilon_{\rm BCG}>$0.15 have their masses divided by
0.93.  The LM relation corrected for triaxiality, Eddington bias and
centroid position is given by the dashed-dotted cyan line in
Figure~\ref{fig:wlLM_bias}.  

By calculating $\epsilon_{\rm BCG}$ for each of our clusters, we have
found that 60$\%$ appear to be viewed near the LOS, suggesting
that weak lensing surveys preferentially select clusters elongated
along the LOS.  Comparing to \cite{2012ApJ...754..119M}, an
X-ray selected cluster sample, only $\sim$18$\%$ of their clusters
have $\epsilon_{\rm BCG}\le$0.15.    

\subsubsection{Weak Lensing Masses within X-ray $r_{500}$}
\label{sec:r500corr}

\begin{figure}
\begin{center}
\includegraphics[clip=true, width=6.0cm, angle=270]{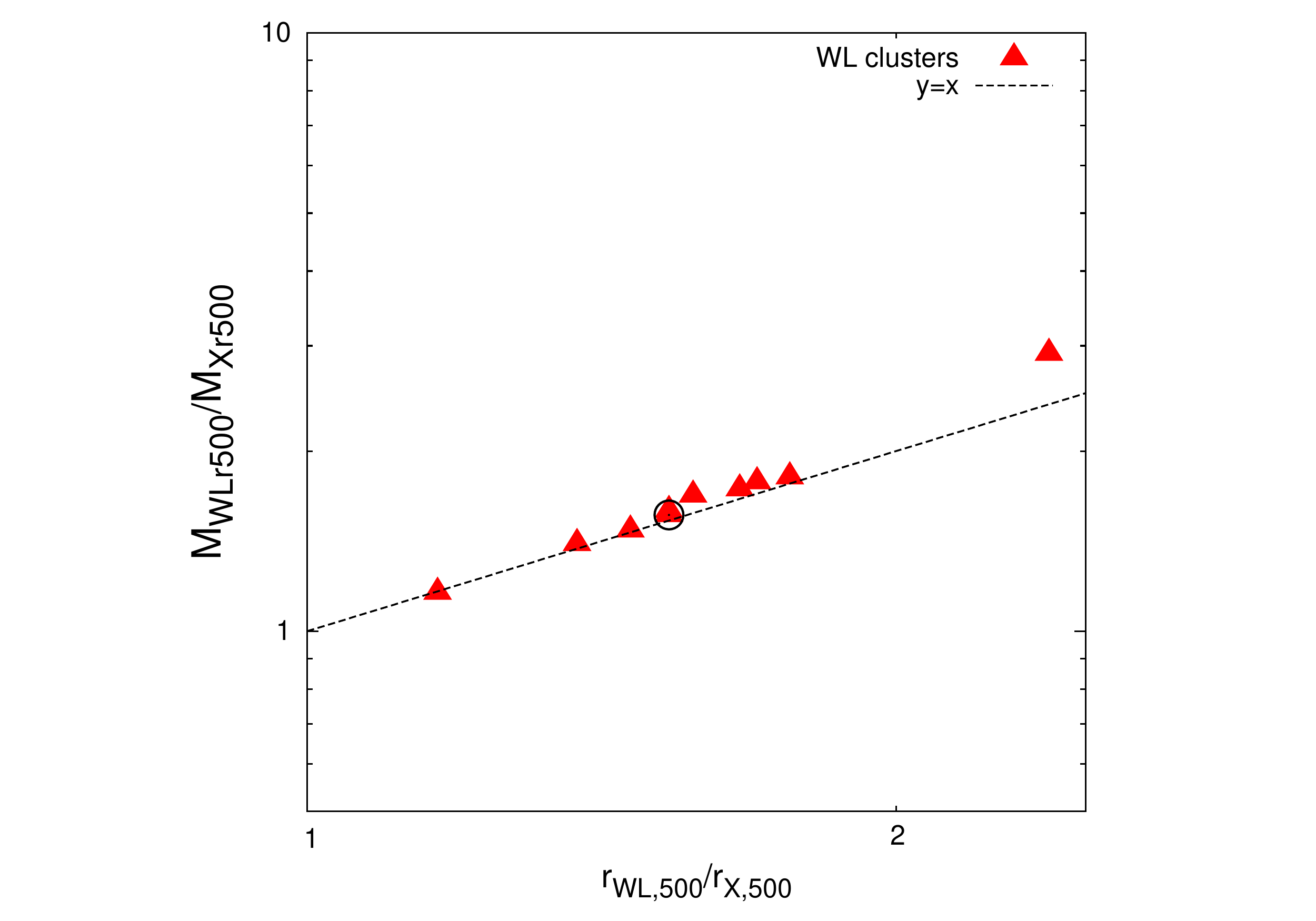}
\end{center}
\caption[]{\small{Plots the ratio of the WL mass derived within the WL
    $r_{WL,500}$ and X-ray $r_{X,500}$, against the ratio of 
    $r_{WL,500}$/$r_{X,500}$.  The cluster SLJ1634 is highlighted by the
    black circle.}\label{fig:r500offset}}
\end{figure}

As stated in $\S$~\ref{sec:wlcentroid}, the original WL analysis was
independent of the X-ray analysis.  The LM relation shown in
Figure~\ref{fig:wlLM} had the masses derived within a WL derived $r_{500}$
($r^{\rm WL}_{500}$) and the luminosities calculated within an X-ray
derived $r_{500}$ ($r^{\rm X}_{500}$).  In order to compare consistently
with the \cite{2010MNRAS.406.1759M} LM relation, whose masses were
derived within $r^{\rm X}_{500}$, we recalculate the WL masses within
$r^{\rm X}_{500}$ for each cluster.  Figure~\ref{fig:r500offset} plots
the change in WL mass against the ratio of the WL and X-ray derived
$r_{500}$.  The WL masses decrease on average by a factor of
$\approx$1.6.  The correction for centroid position, Eddington bias,
triaxiality and the calculation of the WL mass within 
$r^{\rm X}_{500}$, is given by the red solid line in
Figure~\ref{fig:wlLM_bias}.

From Figure~\ref{fig:wlLM_bias} we can clearly see the largest
correction to the WL masses is when they are determined within X-ray
derived $r_{500}$ values.  However, this partially obscures the
true discrepancy.  This shows that the agreement between WL and X-ray
masses is improved if the same aperture is used for both, but this is
no longer a WL $M_{500}$.  We investigate this further in
Section~\ref{sec:wlmasscomp}. 

\subsubsection{Corrected Luminosity-Mass Relation}
\label{sec:wlLM_bias}

\begin{table}
\caption[]{\small{Table of corrections applied to the WL derived mass
    for each cluster.}\label{tab:corrections}} 
\begin{center}
\begin{tabular}{ccccc}
\hline\hline
 & \multicolumn{4}{c}{Correction} \\
\hline
Cluster & Centroid & Edd. Bias & Ellipticity & r$_{500}$ \\ 
\hline 
SLJ0225.7--0312 & 1.05 & 1.20 & 1.17 & 1.16 \\
SLJ1647.7+3455 & 0.93 & 1.34 & 1.17 & 1.73 \\
\hline
SLJ0850.5+4512 & 1.10 & 1.34 & 1.17 & 1.80 \\
SLJ1000.7+0137 & 1.26 & 1.12 & 0.93 & 1.68 \\
SLJ1135.6+3009 & 1.28 & 1.12 & 0.93 & 1.78 \\
SLJ1204.4--0351 & 0.94 & 1.54 & 0.93 & 1.40 \\
SLJ1335.7+3731 & 1.10 & 1.34 & 1.17 & 2.92 \\
SLJ1337.7+3800 & -- & 1.28 & 0.93 & 1.48 \\  
SLJ1602.8+4335 & 0.98 & 1.18 & 1.17 & 1.59 \\
SLJ1634.1+5639 & -- & -- & -- & -- \\
\hline
\end{tabular}
\end{center}
\end{table}

\begin{table}
\begin{center}
\caption[]{\small{Observed scaling relations.  Each scaling relation is
    fit with a power law relation of the form 
    $E(z)^{\alpha}L_{\rm X} = A_{0}(X/X_{0})^{\beta}$, where
    $X_{0}$ is 2 keV and 2$\times$10$^{14}$ M$_{\odot}$ for kT and $M_{\rm X}$
    respectively.}\label{tab:wlscaling}}
\begin{tabular}{lcc}
\hline
 & Normalisation & Slope \\
\hline
$L_{\rm Xc}-kT_{\rm c}$ & $A_{\rm LT_{\rm X,c}}$(10$^{43}$) &
$B_{\rm LT_{\rm X,c}}$ \\
 & 2.4$\pm$0.5 & 2.6$\pm$0.7 \\
\hline
$L_{\rm X,c}-M_{\rm 500,WL}$ & $A_{\rm LM}$(10$^{43}$) &
$B_{\rm LM}$ \\
 & 1.9$\pm$0.2 & 1.33 (fixed) \\
\hline
\multicolumn{3}{l}{Centroid Corrected} \\
$L_{\rm X,c}-M_{\rm 500,WL}$ & $A_{\rm LM}$(10$^{43}$) &
$B_{\rm LM}$ \\
 & 2.2$\pm$0.3 & 1.33 (fixed) \\
\hline
\multicolumn{3}{l}{Centroid and Edd. Bias Corrected} \\
$L_{\rm X,c}-M_{\rm 500,WL}$ & $A_{\rm LM}$(10$^{43}$) &
$B_{\rm LM}$ \\
 & 2.9$\pm$0.5 & 1.33 (fixed) \\
\hline
\multicolumn{3}{l}{Centroid, Edd. Bias and Ellipticity Corrected} \\
$L_{\rm X,c}-M_{\rm 500,WL}$ & $A_{\rm LM}$(10$^{43}$) &
$B_{\rm LM}$ \\
 & 3.0$\pm$0.5 & 1.33 (fixed) \\
\hline
\multicolumn{3}{l}{Centroid, Edd. Bias, Ellipticity and r$_{500}$ Corrected} \\
$L_{\rm X,c}-M_{\rm 500,WL}$ & $A_{\rm LM,corr}$(10$^{43}$) &
$B_{\rm LM,corr}$ \\
 & 5.6$\pm$0.9 & 1.33 (fixed) \\
\hline
\end{tabular}
\end{center}
\end{table}

\begin{figure}
\begin{center}
\includegraphics[clip=true, width=6.0cm, angle=270]{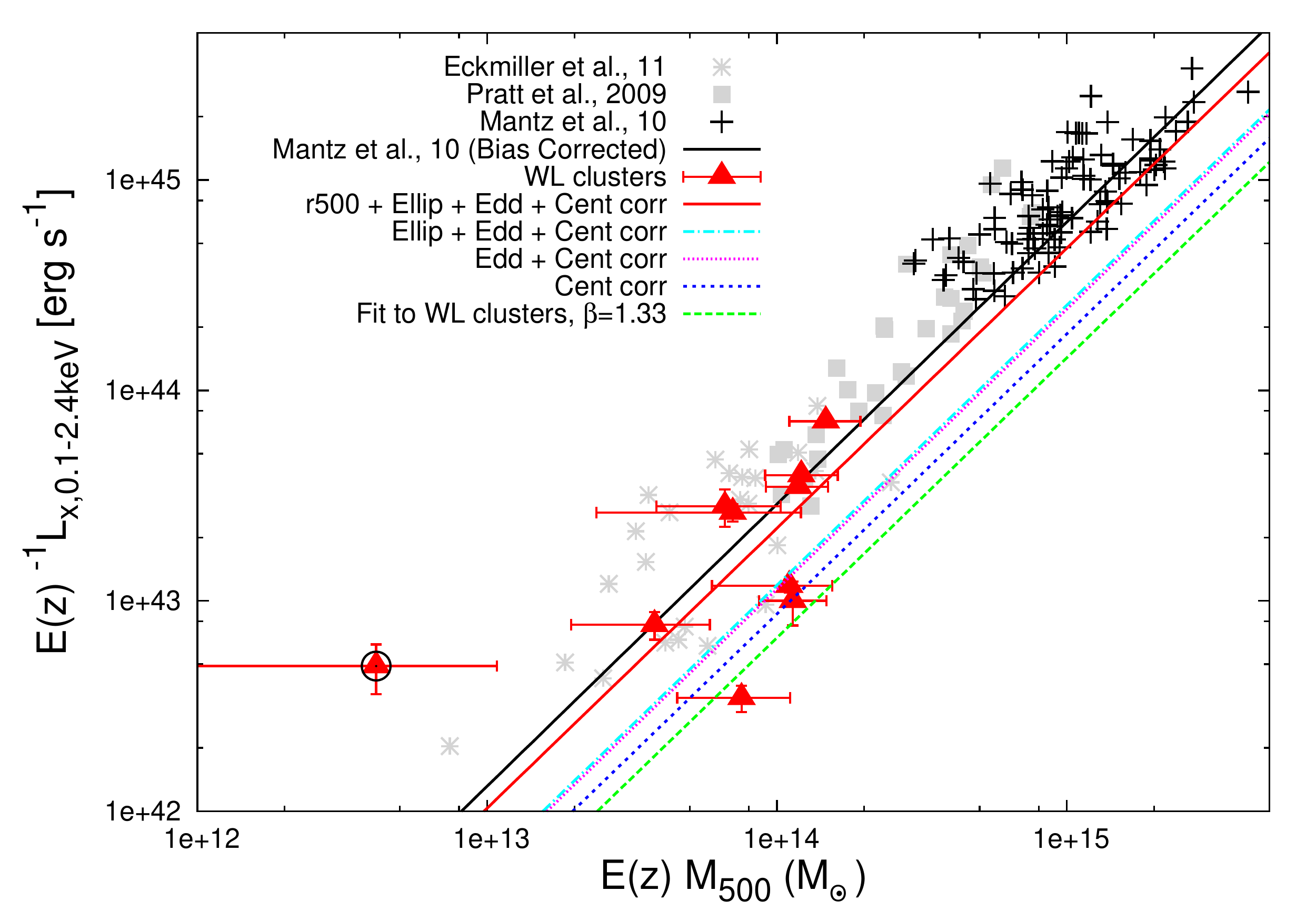}
\end{center}
\caption[]{\small{Figure showing the LM relation for the WL slected
    clusters (red triangles), with the WL masses corrected for
    centroid position ($\S$~\ref{sec:wlcentroid}), Eddington bias
    ($\S$~\ref{sec:eddbias}) and triaxiality
    ($\S$~\ref{sec:wlellipt}).  Finally, the WL masses are derived
    within the X-ray determined $r_{\rm 500}$ (see
    $\S$~\ref{sec:r500corr}).  The cluster SLJ1634 is highlighted by the
    black circle.  A power law fit for the LM relation with
    the masses corrected for these effects is given by the red solid
    line. Once again compared to the clusters given in 
    \cite{2010MNRAS.406.1759M}(black crosses), and corresponding bias
    corrected LM relation (black line).  The clusters from
    \cite{2009A&A...498..361P} and groups from
    \cite{2011A&A...535A.105E} are also plotted, but shaded grey for
    clarity.} \label{fig:wlLM_bias}}
\end{figure}    

In
Sections~\ref{sec:wlcentroid},~\ref{sec:eddbias},~\ref{sec:wlellipt}
and \ref{sec:r500corr},  we presented a series of systematic and
statistical corrections to our WL cluster masses. We re-derive the
LM relation for the WL selected
clusters, with the WL masses corrected for these effects.
Figure~\ref{fig:wlLM_bias} plots the corrected LM relation, with each
of the lines corresponding to a fit with the cumulative effect of
these corrections taken into account. From the final fit to the mass
corrected LM relation (Fig~\ref{fig:wlLM_bias}, red solid line), we find 
$A_{\rm LM,corr}$=(4.7$\pm$0.6)$\times$10$^{43}$ ergs s$^{-1}$, with
the slope fixed at $\beta$=1.33.  The comparison with the
\cite{2010MNRAS.406.1759M} LM relation has been greatly reduced, and
we find that the WL selected clusters are now consistent with X-ray
selected clusters samples, albeit with three clusters being low
luminosity outliers.  We note that the clusters are only consistent
when using WL masses derived within r$^{X}_{500}$, which may obscure
the underlying discrepency.  These corrections therefore show that
clusters detected from WL surveys suffer from biases which have a
large effect on the derived LM relation.  By considering a series
of plausible biases on the WL mass, the discrepancy in the LM
relation, when compared to X-ray selected clusters, can be resolved.     

\subsection{Further Potential Mass Biases}

In the following sections we discuss further biases which may have an
effect when comparing WL and X-ray hydrostatic mass calculations.  

\subsubsection{WL Mass Bias due to Large-Scale Structure}
\label{sec:LSS}

It has been shown that correlated and uncorrelated large-scale
structure (LSS) along the LOS, can both induce bias and
scatter in the WL mass
\citep[e.g][]{1999ApJ...520L...9M,2001A&A...370..743H,2004APh....22...19W,2005NewA...10..676D,2011MNRAS.413..301N,2011ApJ...740...25B}.
We should also note that the effects of LSS and triaxiality are
related, where it can be considered that the major axis of halos is
correlated with filamentary LSS
\citep[e.g.][]{1997ApJ...479..632S,2002A&A...395....1F,2005ApJ...618....1H,2007ApJ...657...30L,2009ApJ...706..747Z,2011MNRAS.413..301N}.
The distribution of BCG ellipticities suggests that 60$\%$ of our
clusters are viewed with the major axis close to the LOS,
from which we can assume that they will also be affected by LSS along
the LOS.  However, \cite{2010A&A...514A..93M} note that since they
consider all the mass within 20 h$^{-1}$Mpc in their lensing
simulations, the effect of correlated LSS is partially included in
their estimate of the lensing mass.  Therefore, the correction applied
to our WL masses in Section~\ref{sec:wlellipt} based upon
\cite{2010A&A...514A..93M}, should partially include corrections for
LSS.  The effect of uncorrelated LSS has been found to not bias the WL
mass, but adds additional scatter $\approx$15-30$\%$ as a function of
cluster mass \citep{2011MNRAS.412.2095H,2011ApJ...740...25B}.      

\subsubsection{Underestimates of X-ray Hydrostatic Mass}
\label{sec:xmassbias}

Simulations have shown that hydrostatic mass estimates of galaxy
clusters are systematically biased towards underestimates of the true
mass
\citep[e.g.][]{2004MNRAS.351..237R,2004MNRAS.355.1091K,2007ApJ...655...98N}.  
Since hydrostatic estimates only take into account thermal
pressure, the presence of additional pressure from random gas
motions are neglected, and hence bias the mass low.  It has been shown
from simulations that up to $\approx$5\%-20\% of pressure support
comes from random gas motions
\citep[e.g.][]{2006MNRAS.369.2013R,2009ApJ...705.1129L}.  However, the
picture from observations is unclear.  A recent study of 22 clusters
detected by {\em Planck}, found that the average X-ray hydrostatic
masses were 0.688$\pm$0.072 times lower than the masses determined
from WL \citep{2014arXiv1402.2670V}. Since
the masses of the \cite{2010MNRAS.406.1759M}
clusters are derived from a relation based upon X-ray hydrostatic mass
estimates, in principle we can assume that their cluster
masses are underestimated at the 10s\% level.  

As stated in Section~\ref{sec:wlLM}, the masses in
\cite{2010MNRAS.406.1759M} are derived from the gas mass.  They note
that they incorporate a systematic fractional bias at $r_{500}$ of
0.0325$\pm$0.06 on the gas mass measurements, based upon the
simulations of \cite{2007ApJ...655...98N}.  Therefore, this bias will
be incorporated in the LM relation comparison.   However, as our
clusters probe a lower mass distribution than the
\cite{2010MNRAS.406.1759M} sample (all
$M_{500}>$3$\times$10$^{14}M_{\odot}$), could the offset be
explained by an increasing hydrostatic mass bias for lower cluster
masses?  The effect would be to not only lower the normalisation of
the X-ray LM relation, but to also increase the slope, further
decreasing the discrepancy of the LM
relations. However, our comparison to the lower mass clusters of
\cite{2009A&A...498..361P} and groups of \cite{2011A&A...535A.105E} do
not offer better agreement with our WL selected clusters.  This has also
been investigated based on simulations.  \cite{2008ApJ...681..167J} showed that
there is a weak correlation of mass underestimate with cluster mass,
with low mass clusters ($\approx$10$^{14}$M$_{\odot}$) having larger
mass bias.
\cite{2008ApJ...681..167J} find an underestimate of
$\sim$20\% at 10$^{14}$M$_{\odot}$.  Recently,
\cite{2014arXiv1402.3267I} compared X-ray and WL cluster masses for a
sample of 8 clusters in the 0.4$\leq$z$\leq$0.5 redshift range.  They
find a 1:1 correlation between the two mass estimates, but observe an
intriguing mass bias at low WL masses at the $\sim$2$\sigma$ level.
However, other studies have found no hydrostatic mass bias
as a function of mass \citep[e.g.][]{2009ApJ...705.1129L}.

\subsection{Comparison to X-ray Selected Cluster Sample with Weak
  Lensing Based Masses}
\label{sec:wlmasscomp}

\begin{figure}
\begin{center}
\includegraphics[clip=true, width=5.9cm, angle=270]{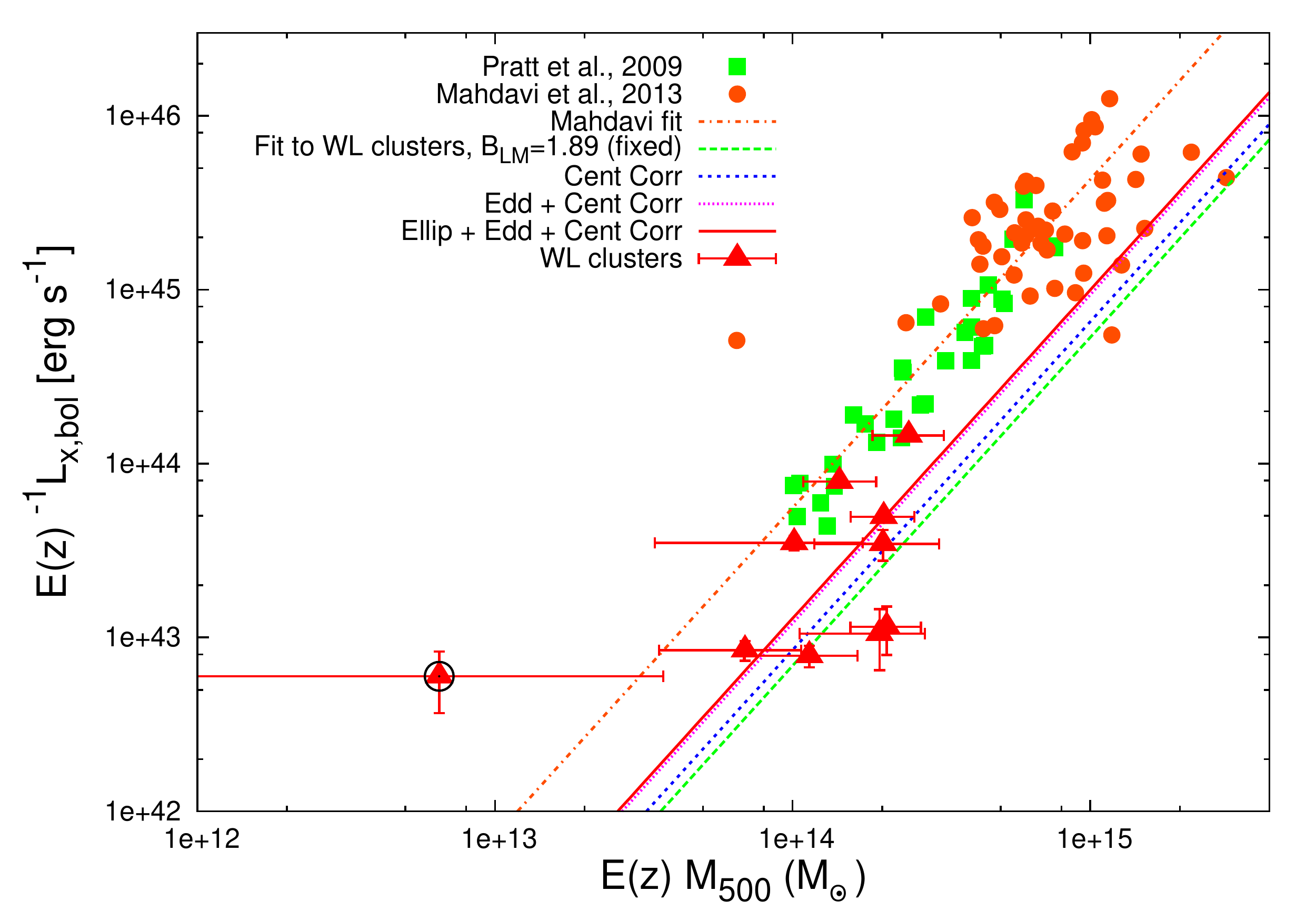}
\end{center}
\caption[]{\small{Figure showing the LM relation for the WL selected
    clusters (red triangles), with the WL masses corrected for
    centroid position ($\S$~\ref{sec:wlcentroid}), Eddington bias
    ($\S$~\ref{sec:eddbias}) and triaxiality
    ($\S$~\ref{sec:wlellipt}).  The cluster SLJ1634 is highlighted by the
    black circle.  A power law fit for the LM relation with
    the masses corrected for these effects is given by the red solid
    line. Also plotted are the \citet[][brown
      circles]{2013ApJ...767..116M} clusters with WL masses in 
   $r^{\rm WL}_{500}$.  For reference, the \citet[][green
      squares]{2009A&A...498..361P} clusters are plotted, however
    these are X-ray mass estimates.} \label{fig:wlLM_mahdavicomp}}
\end{figure}     

We compare to the sample of clusters studied in
\cite{2013ApJ...767..116M}, which is an
X-ray selected cluster sample with WL derived cluster masses.  The 
main advantage of choosing this sample is that the WL masses are
derived within a WL determined $r_{500}$. As we have shown in
Figure~\ref{fig:wlLM_bias}, the choice of $r_{500}$ had the largest
effect on the derived WL mass and may obscure any further underlying 
biases.  By making this comparison we can remove the large effect of
the choice of $r_{500}$.  Furthermore, by comparing to WL
masses the effects of hydrostatic mass bias (see
$\S$~\ref{sec:xmassbias}), and to a lesser extent the WL mass bias due 
to LSS (see $\S$~\ref{sec:LSS}, but could still be larger in our
sample compared to \cite{2013ApJ...767..116M}, due to the WL
selection) will be reduced or removed.

Figure~\ref{fig:wlLM_mahdavicomp} plots the LM relation for our WL
selected clusters (red triangles), and compared to the clusters
studied in \citet[][brown circles]{2013ApJ...767..116M}\footnote{The
  data were downloaded from http://sfstar.sfsu.edu/cccp/}.  For the
purposes of this comparison the X-ray luminosities are bolometric and
in each case the WL masses are determined within a WL defined
$r_{500}$. We fit a power law of the form
$L_{X}=E(z)^{2}L_{0}(M/M_{0})^{B_{\rm LM}}$
to the \cite{2013ApJ...767..116M} clusters finding a normalisation of 
$L_{0,M13}$=(2.83$\pm$0.47)$\times$10$^{45}$ergs s$^{-1}$ and slope
$B_{LM,M13}$=1.89$\pm$0.83, with
$M_{0}$=8$\times$10$^{14}$M$_{\odot}$.  We fit to our WL clusters with
the slope fixed at the value found for the \cite{2013ApJ...767..116M}
relation ($B_{LM,M13}$=1.89), and find a normalisation of
$L_{0}$=(3.8$\pm$0.8)$\times$10$^{43}$ergs s$^{-1}$ (at
$M_{0}$=2$\times$10$^{14}$M$_{\odot}$, see red solid line
Fig~\ref{fig:wlLM_mahdavicomp}.  Therefore, for a given mass
we find that our WL selected clusters are 5.4$\pm$1.4 times
under-luminous as compared to the \cite{2013ApJ...767..116M} clusters
in terms of the LM relation, with the masses derived from WL and within
a WL determined $r_{500}$ (corrected for centroid position, Eddington
bias and triaxiality).  Conversely, the masses are over-estimated by
a factor 2.4$\pm$0.3 for a given luminosity.  To validate that the
X-ray selected clusters of \cite{2013ApJ...767..116M}, whose masses
are derived from WL and within r$^{\rm WL}_{500}$, follow other X-ray selected
clusters, we plot the \cite{2009A&A...498..361P} clusters
(Fig~\ref{fig:wlLM_mahdavicomp}, green squares).  Clearly, the two X-ray
selected samples are in agreement, even though the mass estimates are
from two different methods, with our WL selected clusters offset from
both samples. We note that if the \cite{2013ApJ...767..116M} WL
masses are calculated within an X-ray derived $r_{500}$, this has a
negligible effect on their LM relation.  

Since the discrepancy with the \cite{2013ApJ...767..116M} sample is
still prevalent (see Fig~\ref{fig:wlLM_mahdavicomp}), even after
correcting our WL masses for centroid position, Eddington bias and
triaxiality, this implies that there is still some underlying effect
due to the WL selection of these clusters.

\subsection{Are WL Selected Clusters Under-luminous?}
\label{sec:vdisp-lx}

\begin{figure}
\begin{center}
\includegraphics[clip=true, width=5.9cm, angle=270]{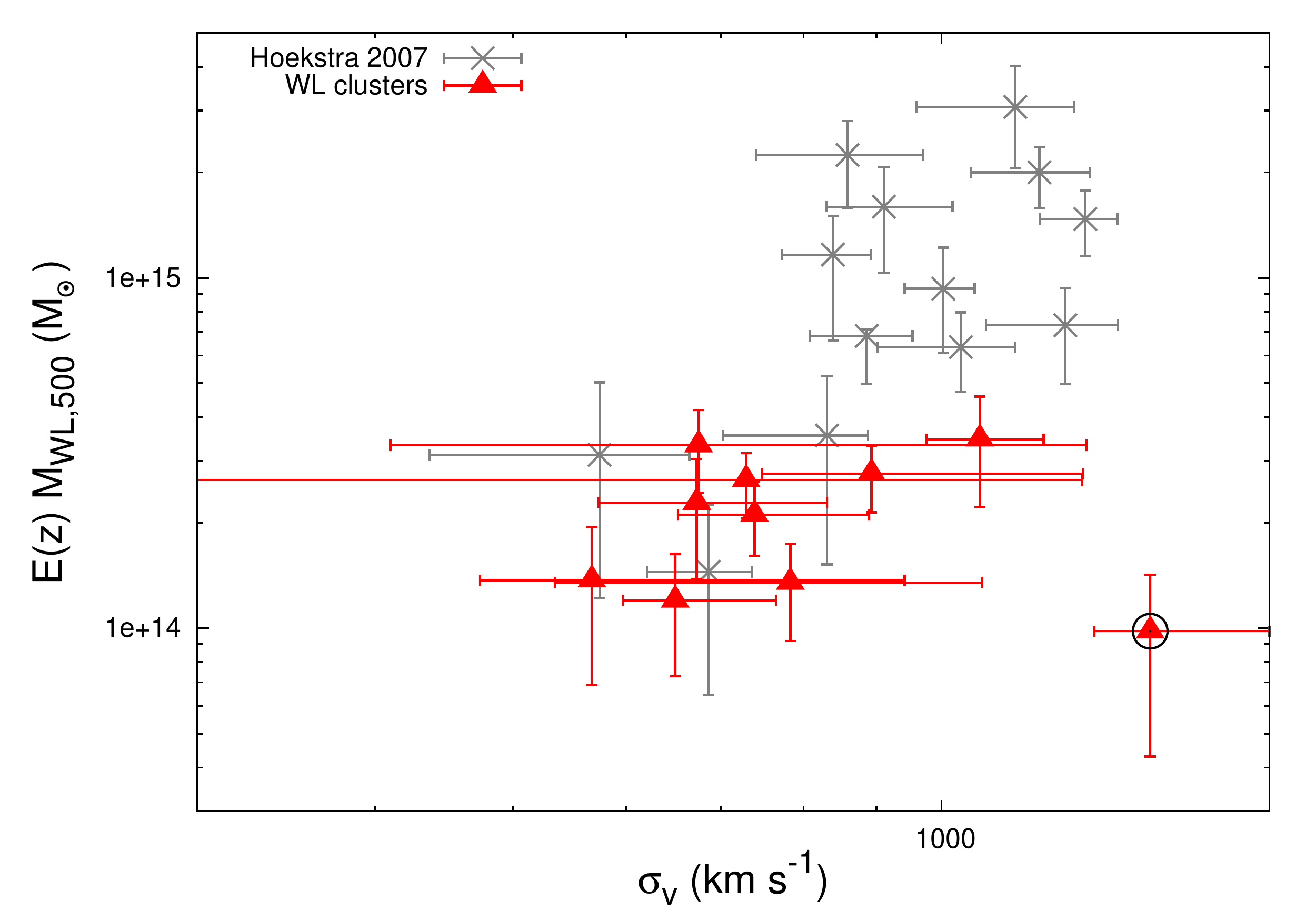}
\end{center}
\caption[]{\small{Figure showing the $M_{\rm WL,500}$-$\sigma_{v}$
    relation for the WL selected clusters (red triangles). The cluster
    SLJ1634 is highlighted by the black circle. The
    clusters studied in \cite{2007MNRAS.379..317H} are plotted for
    comparison (grey crosses).} \label{fig:m500_vdisp}}
\end{figure}

In Section~\ref{sec:systematics} we have shown that by considering various
systematic effects on the WL cluster mass, we are unable to reconcile
the differences seen in the LM relation (see
Figure~\ref{fig:wlLM_mahdavicomp}), without including corrections for
r$_{500}$ which we argue masks the true discrepancy (see
$\S$~\ref{sec:wlmasscomp}, Fig~\ref{fig:wlLM_mahdavicomp}).  Not only
that, our previous work \citep[][Figure 5]{2009PASJ...61..833H} found that 
the WL mass is consistent with previous studies when compared to the
velocity dispersion ($\sigma_{v}$).   We reproduce this comparison
using $M_{\rm WL,500}$ in Figure~\ref{fig:m500_vdisp}.  We find that the
WL clusters (red triangles) are consistent with the X-ray selected
clusters studied in \citet[][grey crosses]{2007MNRAS.379..317H}, in
terms of the $M_{\rm WL,500}$-$\sigma_{v}$ relation.  Therefore, we
are presented with the situation that the WL mass and $\sigma_{v}$ are
consistent and the luminosity and temperature are consistent, implying
the ICM properties are inconsistent with $\sigma_{v}$.

\begin{figure}
\begin{center}
\includegraphics[clip=true, width=5.9cm, angle=270]{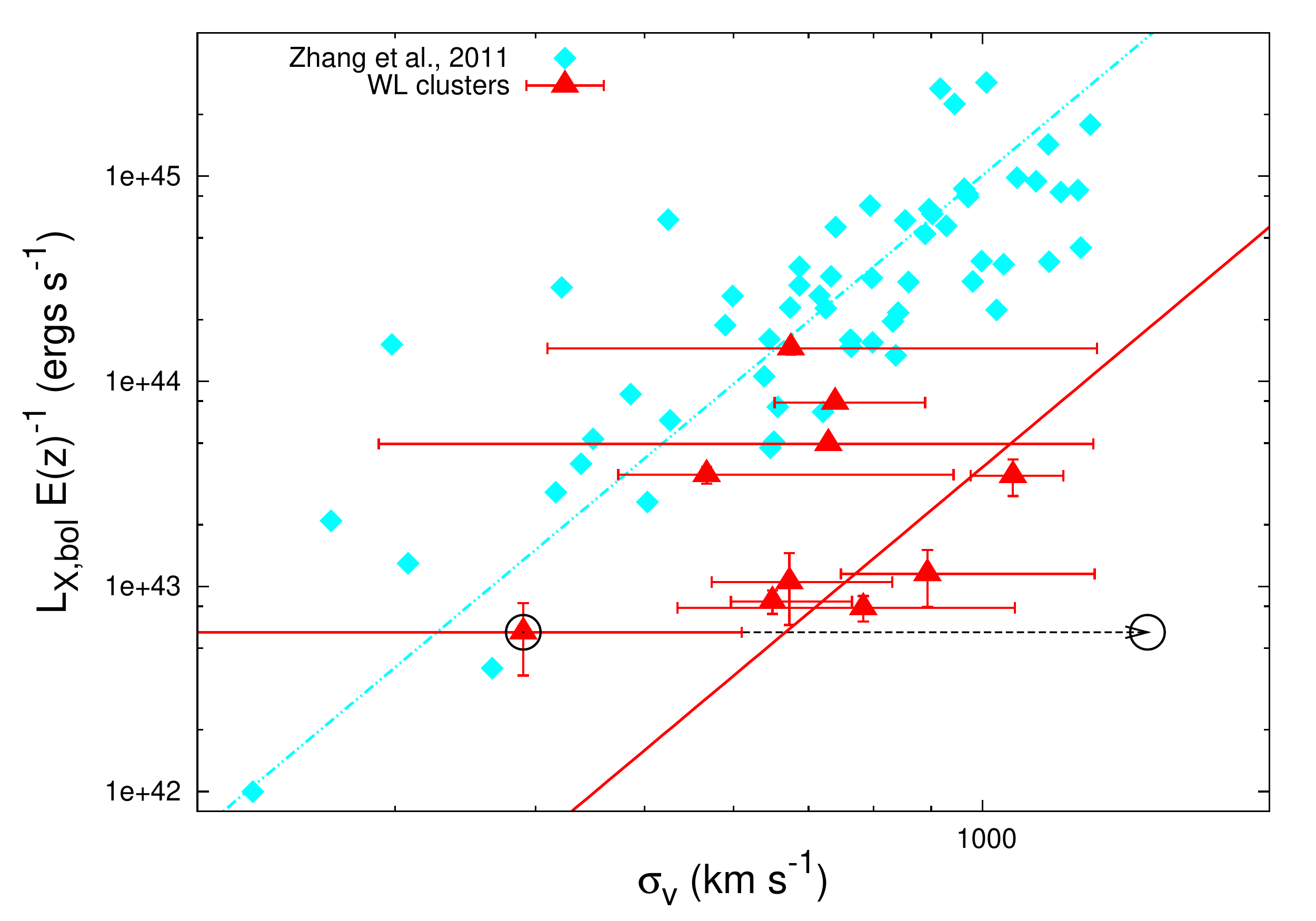}
\end{center}
\caption[]{\small{Figure showing the $L_{\rm X,bol}$-$\sigma_{v}$ for the
    WL selected clusters (red triangles).  The cluster SLJ1634 is
    highlighted by the black circle.  The open black circle represents
    the estimate of the velocity dispersion for SLJ1634 from
    \cite{2009PASJ...61..833H}, when including all spectroscopic
    galaxies. The HIFLUGCS clusters studied in
    \citet[][cyan diamonds]{2011A&A...526A.105Z} is plotted for
    comparison.} \label{fig:vdisp_lx}}
\end{figure}

To quantify this, we derive a $L_{\rm X,bol}$-$\sigma_{v}$ relation for
the WL clusters.  Figure~\ref{fig:vdisp_lx} plots the
$L_{\rm X,bol}$-$\sigma_{v}$ relation for our clusters (red triangles),
compared to the relation for the HIFLUGCS sample of clusters (cyan
diamonds) studied in \cite{2011A&A...526A.105Z}.  The cluster SLJ1634
is highlighted by the black circle.  Due to the composite nature of
this cluster (see $\S$\ref{sec:wlslj1634}), the value of $\sigma_{v}$
from \citet[][$\sigma_{v}$=1402$^{+334}_{-121}$ km s$^{-1}$]{2009PASJ...61..833H}
is likely an overestimate.  We therefore re-derive $\sigma_{v}$ from
our tree analysis using the galaxies from group B (see
Fig~\ref{fig:slj1634}, balck squares), as two of these galaxies are
associated with the X-ray emission, finding 
$\sigma_{v}$=390$\pm$220 km s$^{-1}$. We find that five of the WL
selected clusters are clearly under-luminous for their velocity dispersion.  

These results imply that there is a strong systematic effect remaining
that is influencing the WL and dynamical properties, the ICM
properties, or both.  We consider the most plausible effect to be a
high incidence of projected filamentary structure along the
LOS to these WL selected clusters, leading to enhanced WL
masses and $\sigma_{v}$ relative to halo mass.  From the spectroscopic
follow-up, many of the clusters show signs of non-gaussianity in
their velocity dispersion histograms 
\citep[see Fig 1 in][]{2009PASJ...61..833H}, which would be consistent
with projected filament structure.  This however is based on
relatively small numbers of spectra in each case (an average of 14
galaxies per cluster), so a more detailed analysis is unable to be
performed.         

\cite{1997MNRAS.291..353B} investigated a sample of optically
selected, X-ray under-luminous clusters and proposed
two scenarios for their observed offset in the $L_{\rm X}$-$\sigma_{v}$
relation.  Either the clusters have not yet formed or are at a stage of rapid
mass accretion, which would be manifested in a low X-ray luminosity.
The second scenario is that the clusters are embedded
in filaments viewed along the LOS, which would lead to an
over-estimate of $\sigma_{v}$ due to contamination of galaxies along
the filament.  Other previous studies looking at X-ray under-luminous
clusters \cite[e.g.][]{2007A&A...461..397P,2011MNRAS.412..947B,2011A&A...530A..27C}
favour the scenario that these clusters are still in a stage of
formation or that the gas has been expelled.  However, since we have
measured temperatures for the majority of our clusters
and shown that the gas properties scale consistently in the
luminosity-temperature relation (and consistent to X-ray selected
cluster samples), we favour the scenario that the WL selected clusters
are embedded in filaments viewed near the LOS.  This is further
corroborated by our measurements of the BCG ellipticity, where the
majority of our cluster BCGs appear circular, implying that the WL
selection strongly favours clusters elongated along the LOS
and embedded in correlated filaments.

\section{Summary and Conclusions}
\label{sec:wlconclusion}

We have presented the results of X-ray observations of 10 WL
selected galaxy clusters in order to determine their X-ray
properties.  Our main results are as follows.
\begin{enumerate}
\item We find that the scaling relation between X-ray luminosity and
  temperature is consistent with samples of X-ray selected clusters.  
\item We find that the WL selected clusters are offset from X-ray
  selected clusters and groups in the luminosity-temperature relation,
  implying either the X-ray luminosity is under-estimated or the WL
  mass is over-estimated. 
\item The centroid shifts for the WL clusters show that the majority
  appears dynamically unrelaxed compared to X-ray selected samples,
  suggesting X-ray cluster surveys preferentially detect clusters that
  are morphologically relaxed and/or WL selection favours
  morphologically disturbed clusters.  
\item The cuspiness of the gas density profiles shows that two of the
  clusters in our sample appear to host a cool core.
\item Measuring the ellipticities of the BCGs for each cluster, and
  assuming that they are intrinsically prolate, we
  find that 60$\%$ of the clusters appear to be viewed with their
  major axis close to the line of sight.
\item Through a series of corrections to the WL cluster mass including
  the centroid position, Eddington bias, and triaxiality, and
  calculating the WL masses with an X-ray derived r$_{500}$, we find
  the WL selected clusters are consistent in the LM relation with
  X-ray selected clusters.
\item Comparisons to X-ray selected samples with WL derived
  cluster masses, and within WL defined r$_{500}$s, show
  that the offset in the LM relation is still present, even when
  correcting for centroid position, Eddington bias, and triaxiality.    
\end{enumerate}          

Our results show that WL selected clusters are affected by biases
which, when combined, has a large effect on the calculated WL mass.
We have presented a series of possible biases to explain the large
offset of the LM relation when compared to X-ray selected cluster
samples.  When considering the cumulative effect of these biases on
the WL mass we reconcile the difference in the LM relation.  Our
comparison to an X-ray selected sample of clusters with WL derived
cluster masses, show that even after taking into account biases which
may arise due to the WL selection, there is still a discrepancy in the
LM relation.  This implies that there maybe further biases in the WL
selection, or that the WL selected clusters are embedded in
filamentary structure viewed along the line of sight.

\section*{Acknowledgments}
We thank Massimo Meneghetti for providing data in electronic format of
his simulated clusters.  PAG also acknowledges support from the UK
Science and Technology Facilities Council.  RM is supported by a Royal
Society University Research Fellowship, and European Research Council
grant MIRG-CT-208994.  This research made use of
observations taken with {\em Chandra} awarded in proposal ID 12800522
(P.I. PAG), data from the {\em Chandra} data archive and data from the
{\em XMM-Newton} data archive.

\onecolumn
\makeatletter
\def\fps@figure{h}
\makeatother
\appendix
\begin{center}
\section{X-ray and {\em Subaru} Images of Individual Clusters}
\end{center}
\label{sec:wlimages}

\begin{figure*}
\begin{center}
\setlength{\unitlength}{1in}
\begin{picture}(6.4,3.0)
\put(0.0,0.06){\scalebox{0.40}{\includegraphics[clip=true]{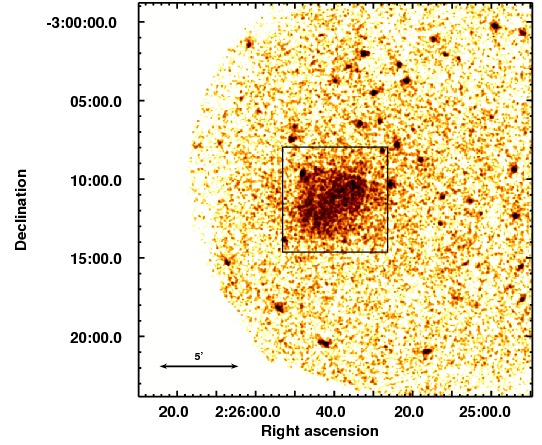}}}  
\put(3.50,0.1){\scalebox{0.37}{\includegraphics[clip=true]{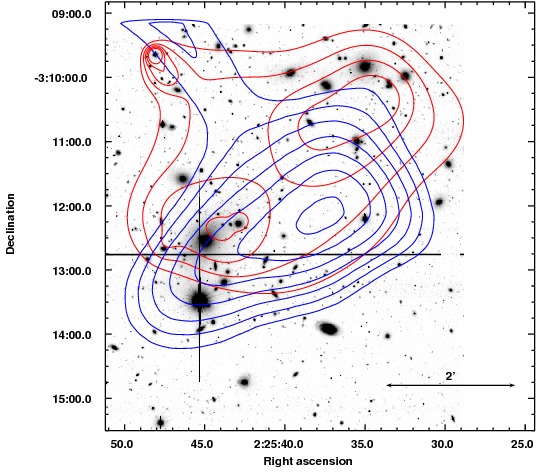}}}
\put(1.7,-0.05){(a)}
\put(5.0,-0.05){(b)}
\end{picture}
\end{center}
\caption[]{\small{Images for the cluster SLJ0225 (z=0.1395).  Figure (a)
shows the {\em XMM} image of the cluster in the 0.7 -- 2.0 keV band smoothed
by a Gaussian of 1.5 pixel radius (where 1 pixel = 2.2$^{\prime}$);
(b) shows a {\em Subaru} image of the cluster within a region
400$\times$400$^{\prime\prime}$ in length (shown by the black box in
Fig(a)) with X-ray contours (red) and WL mass contours
(blue) over-plotted.  The X-ray contours were constructed from an
adaptively smoothed image of the total MOS data, due to the cluster
falling on the chip gaps of the pn-camera.  The WL contours are taken from
\cite{2009PASJ...61..833H}.}\label{fig:slj0225}}
\end{figure*}

\begin{figure*}
\begin{center}
\setlength{\unitlength}{1in}
\begin{picture}(6.4,3.0)
\put(0.07,0.08){\scalebox{0.38}{\includegraphics[clip=true]{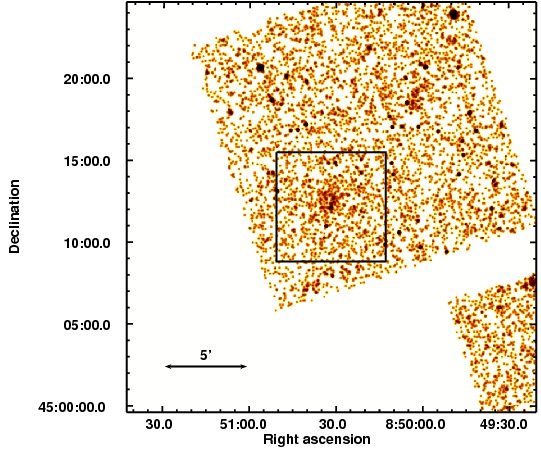}}}  
\put(3.50,0.09){\scalebox{0.36}{\includegraphics[clip=true]{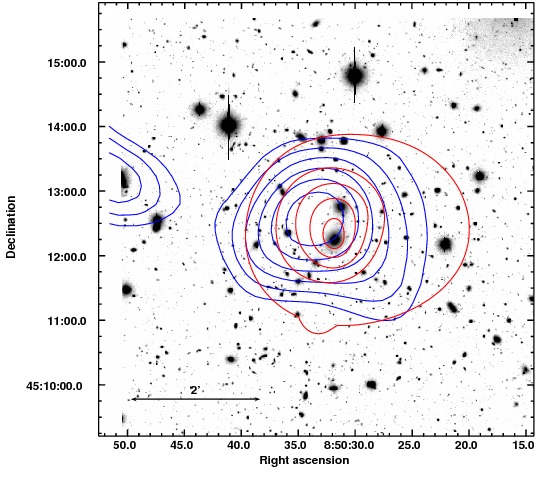}}}
\put(1.7,-0.05){(a)}
\put(5.0,-0.05){(b)}
\end{picture}
\end{center}
\caption[]{\small{Images for the cluster SLJ0850 (z=0.1938). (a) shows
    a {\em Chandra} image in the 0.7--2.0 keV band within a region
    25$\times$25$^{\prime}$ in length, the image is smoothed by a
    Gaussian of 1.5 pixel radius (where 1 pixel =
    1.968$^{\prime\prime}$); (b) shows a {\em Subaru} image of the cluster
    within a region 400$\times$400$^{\prime\prime}$ in length (shown
    by the black box in Fig (a)) with the X-ray contours (red) and
    WL mass contours (blue) over-plotted.  The X-ray
    contours were constructed from an adaptively smoothed {\em
      Chandra} image of the cluster.} \label{fig:slj0850}}
\end{figure*}

\begin{figure*}
\begin{center}
\setlength{\unitlength}{1in}
\begin{picture}(6.4,3.0)
\put(0.12,0.08){\scalebox{0.40}{\includegraphics[clip=true]{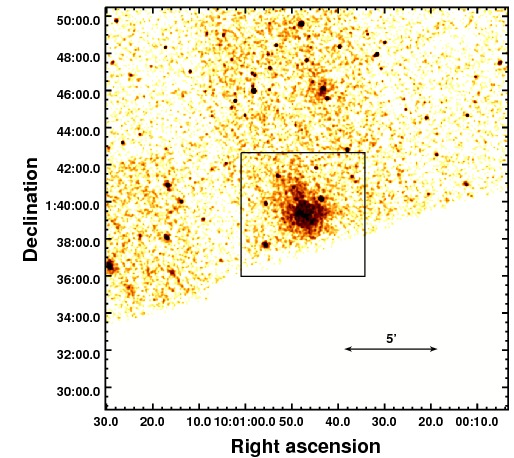}}}  
\put(3.50,0.10){\scalebox{0.37}{\includegraphics[clip=true]{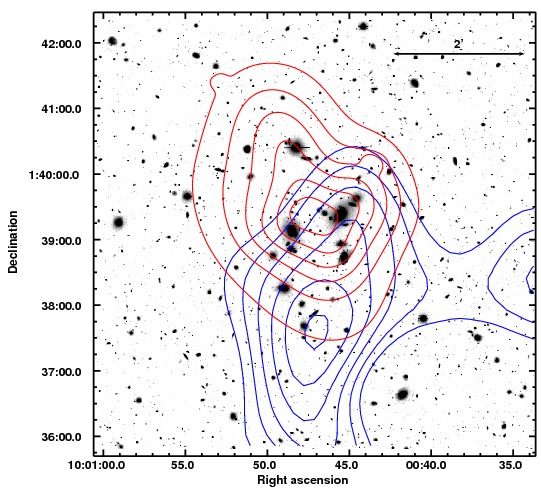}}}
\put(1.7,-0.05){(a)}
\put(5.0,-0.05){(b)}
\end{picture}
\end{center}
\caption{\small{Images for the cluster SLJ1000 (z=0.2166). Figures (a)
    and (b) same as \ref{fig:slj0850}.} \label{fig:slj1000}}
\end{figure*}

\begin{figure*}
\begin{center}
\setlength{\unitlength}{1in}
\begin{picture}(6.4,3.0)
\put(0.07,0.08){\scalebox{0.39}{\includegraphics[clip=true]{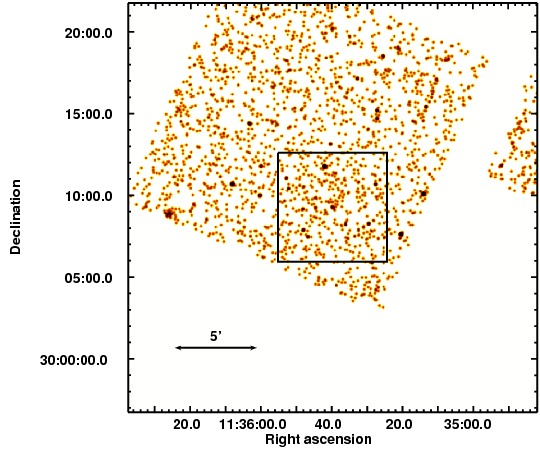}}}  
\put(3.50,0.11){\scalebox{0.475}{\includegraphics[clip=true]{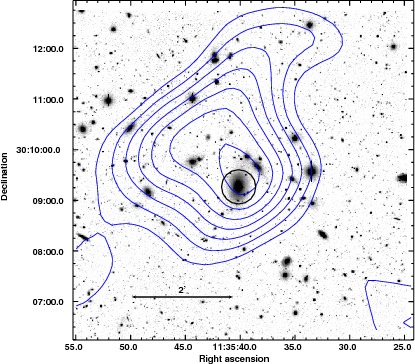}}}
\put(1.7,-0.05){(a)}
\put(5.0,-0.05){(b)}
\end{picture}
\end{center}
\caption{\small{Images for the cluster SLJ1135 (z=0.2078). Figures (a)
    and (b) same as \ref{fig:slj0850}, however due to the low SNR of
    this cluster an adaptively smoothed image of the cluster was not
    obtained.} \label{fig:slj1135}}
\end{figure*}

\begin{figure*}
\begin{center}
\setlength{\unitlength}{1in}
\begin{picture}(6.4,3.0)
\put(0.07,0.08){\scalebox{0.38}{\includegraphics[clip=true]{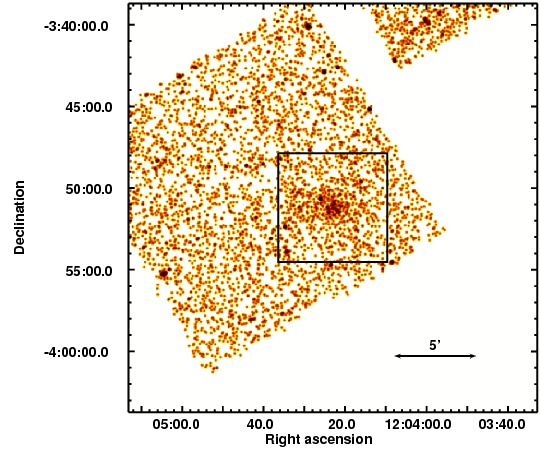}}}  
\put(3.50,0.09){\scalebox{0.36}{\includegraphics[clip=true]{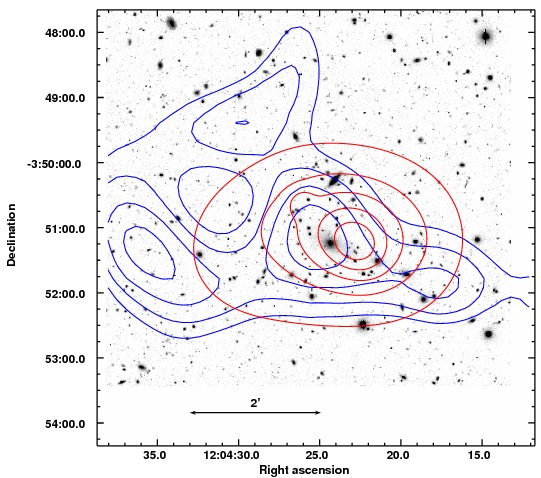}}}
\put(1.7,-0.05){(a)}
\put(5.0,-0.05){(b)}
\end{picture}
\end{center}
\caption{\small{Images for the cluster SLJ1204 (z=0.2609). Figures (a)
    and (b) same as \ref{fig:slj0850}.} \label{fig:slj1204}}
\end{figure*}

\begin{figure*}
\begin{center}
\setlength{\unitlength}{1in}
\begin{picture}(6.4,3.0)
\put(0.07,0.08){\scalebox{0.38}{\includegraphics[clip=true]{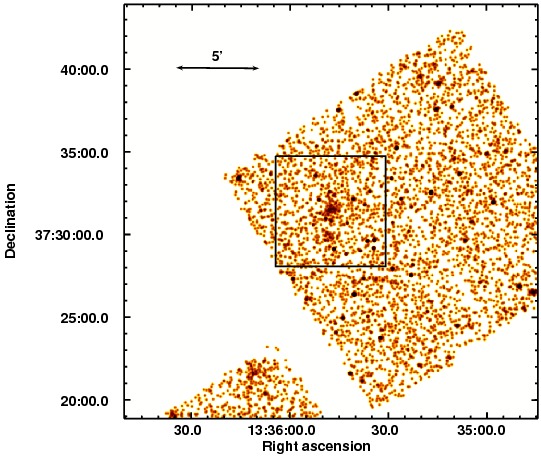}}}  
\put(3.45,0.06){\scalebox{0.368}{\includegraphics[clip=true]{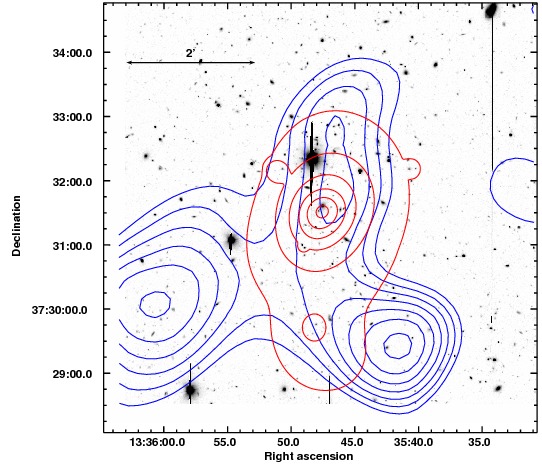}}}
\put(1.7,-0.05){(a)}
\put(5.0,-0.05){(b)}
\end{picture}
\end{center}
\caption{\small{Images for the cluster SLJ1335 (z=0.4070). Figures (a)
    and (b) same as \ref{fig:slj0850}.} \label{fig:slj1335}}
\end{figure*}
  
\begin{figure*}
\begin{center}
\setlength{\unitlength}{1in}
\begin{picture}(6.4,3.0)
\put(0.07,0.08){\scalebox{0.38}{\includegraphics[clip=true]{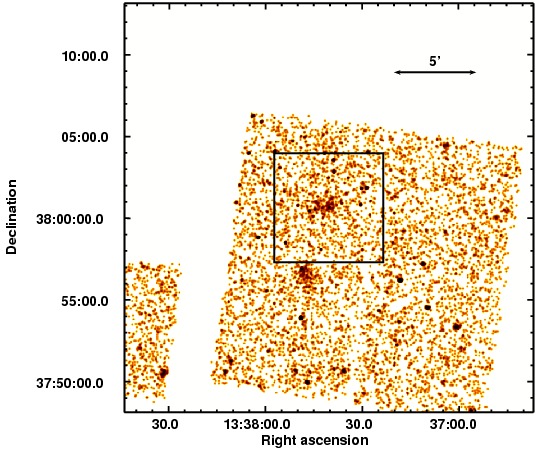}}}  
\put(3.50,0.12){\scalebox{0.36}{\includegraphics[clip=true]{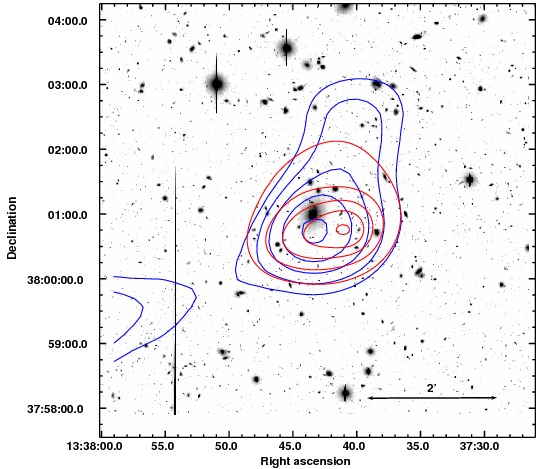}}}
\put(1.7,-0.05){(a)}
\put(5.0,-0.05){(b)}
\end{picture}
\end{center}
\caption{\small{Images for the cluster SLJ1337 (z=0.1798). Figures (a)
    and (b) same as \ref{fig:slj0850}.} \label{fig:slj1337}}
\end{figure*}

\begin{figure*}
\begin{center}
\setlength{\unitlength}{1in}
\begin{picture}(6.4,3.0)
\put(0.07,0.08){\scalebox{0.38}{\includegraphics[clip=true]{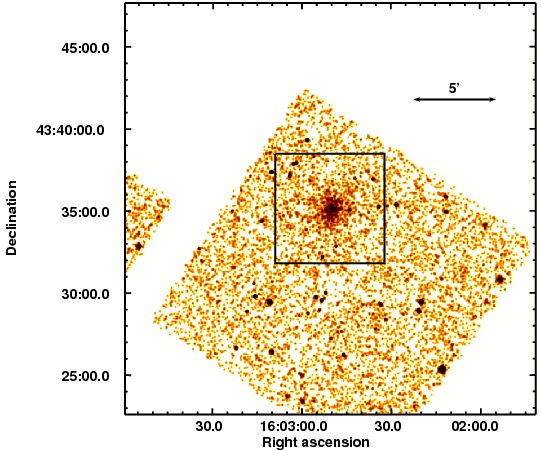}}}  
\put(3.50,0.12){\scalebox{0.36}{\includegraphics[clip=true]{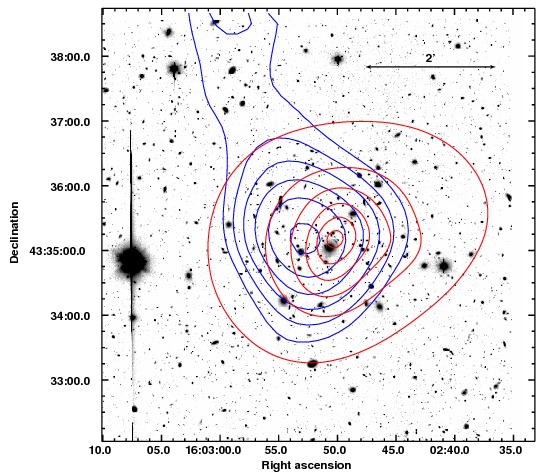}}}
\put(1.7,-0.05){(a)}
\put(5.0,-0.05){(b)}
\end{picture}
\end{center}
\caption{\small{Images for the cluster SLJ1602 (z=0.4155). Figures (a)
    and (b) same as \ref{fig:slj0850}.} \label{fig:slj1602}}
\end{figure*}

\begin{figure*}
\begin{center}
\setlength{\unitlength}{1in}
\begin{picture}(6.4,3.0)
\put(0.07,0.08){\scalebox{0.38}{\includegraphics[clip=true]{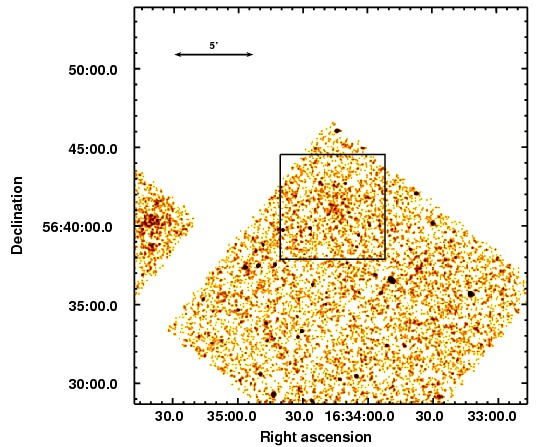}}}  
\put(3.50,0.12){\scalebox{0.36}{\includegraphics[clip=true]{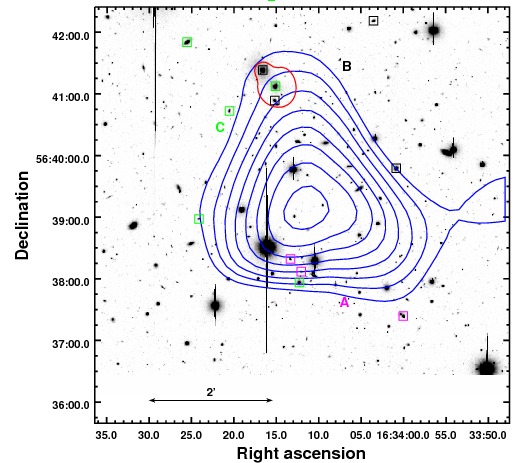}}}
\put(1.7,-0.05){(a)}
\put(5.0,-0.05){(b)}
\end{picture}
\end{center}
\caption{\small{Images for the cluster SLJ1634 (z=0.2377). Figures (a)
    and (b) same as \ref{fig:slj0850}.  Also highlighted on Figure (b)
  is the locations of spectroscopically confirmed cluster galaxies,
  separated into three groups A, B and C (magenta, black and green
  squares respectively) found by our tree analysis
  (see $\S$~\ref{sec:wlslj1634}).} \label{fig:slj1634}}
\end{figure*}

\begin{figure*}
\begin{center}
\setlength{\unitlength}{1in}
\begin{picture}(6.4,3.0)
\put(0.07,0.08){\scalebox{0.38}{\includegraphics[clip=true]{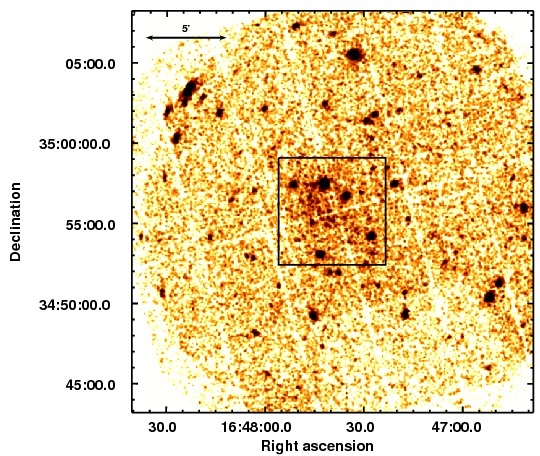}}}  
\put(3.50,0.12){\scalebox{0.37}{\includegraphics[clip=true]{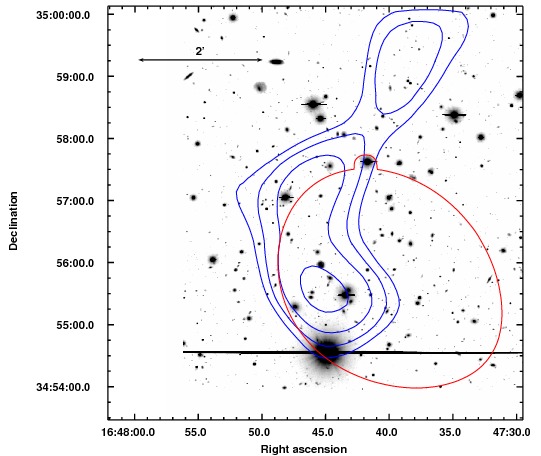}}}
\put(1.7,-0.05){(a)}
\put(5.0,-0.05){(b)}
\end{picture}
\end{center}
\caption{\small{Images for the cluster SLJ1647 (z=0.2592). Figures (a)
    and (b) same as \ref{fig:slj0850}.} \label{fig:slj1647}}
\end{figure*}

\end{document}